\documentclass[11pt,preprint,flushrt]{aastex}
\usepackage{graphicx,amsmath,subfigure,xcolor}
\usepackage{hyperref} 
\hypersetup{
    colorlinks=true, 
    linktoc=all,     
    linkcolor=blue,  
}
\def\eqq#1{Equation~(\ref{#1})}
\def\etal{{\it et al.}}
\def\ie{{\it i.e.}}
\def\eg{{\it e.g.}}

\newcommand{\vpi}{\mbox{\boldmath $\pi$}}
\newcommand{\vx}{\mbox{${\bf x}$}}
\newcommand{\vm}{\mbox{${\bf m}$}}
\newcommand{\vv}{\mbox{${\bf v}$}}
\newcommand{\vk}{\mbox{${\bf k}$}}
\newcommand{\vvr}{\mbox{${\bf r}$}}
\newcommand{\matA}{\mbox{$\mathbf A$}}
\newcommand{\matB}{\mbox{$\mathbf B$}}

\newcommand{\wcsfit}{\textsc{WcsFit}}
\newcommand{\scamp}{\textsc{scamp}}

\newcommand{\edit}[1]{#1}

\usepackage{eso-pic}

\AddToShipoutPictureBG*{%
  \AtPageUpperLeft{%
    \hspace{0.75\paperwidth}%
    \raisebox{-3.5\baselineskip}{%
      \makebox[0pt][l]{\textnormal{DES 2016-0215}}
}}}%

\AddToShipoutPictureBG*{%
  \AtPageUpperLeft{%
    \hspace{0.75\paperwidth}%
    \raisebox{-4.5\baselineskip}{%
      \makebox[0pt][l]{\textnormal{FERMILAB-PUB-17-057-AE}}
}}}%
\slugcomment{Version 1.1, Submitted to PASP}
\begin{document}

\title{Astrometric calibration and performance of the Dark Energy Camera}


\def\andname{}

\author{
G.~M.~Bernstein\altaffilmark{1},
R.~Armstrong\altaffilmark{2},
A.~A.~Plazas\altaffilmark{3},
A.~R.~Walker\altaffilmark{4},
T. M. C.~Abbott\altaffilmark{4},
S.~Allam\altaffilmark{5},
K.~Bechtol\altaffilmark{6},
A.~Benoit-L{\'e}vy\altaffilmark{7,8,9},
D.~Brooks\altaffilmark{8},
D.~L.~Burke\altaffilmark{10,11},
A. Carnero Rosell\altaffilmark{12,13},
M.~Carrasco~Kind\altaffilmark{14,15},
J.~Carretero\altaffilmark{16},
C.~E.~Cunha\altaffilmark{10},
L.~N.~da Costa\altaffilmark{12,13},
D.~L.~DePoy\altaffilmark{17},
S.~Desai\altaffilmark{18},
H.~T.~Diehl\altaffilmark{5},
T.~F.~Eifler\altaffilmark{3},
E.~Fernandez\altaffilmark{16},
P.~Fosalba\altaffilmark{19},
J.~Frieman\altaffilmark{5,20},
J.~Garc\'ia-Bellido\altaffilmark{21},
D.~W.~Gerdes\altaffilmark{22,23},
D.~Gruen\altaffilmark{10,11},
R.~A.~Gruendl\altaffilmark{14,15},
J.~Gschwend\altaffilmark{12,13},
G.~Gutierrez\altaffilmark{5},
K.~Honscheid\altaffilmark{24,25},
D.~J.~James\altaffilmark{26,4},
S.~Kent\altaffilmark{5,20},
E.~Krause\altaffilmark{10},
K.~Kuehn\altaffilmark{27},
N.~Kuropatkin\altaffilmark{5},
T.~S.~Li\altaffilmark{5,17},
M.~A.~G.~Maia\altaffilmark{12,13},
M.~March\altaffilmark{1},
J.~L.~Marshall\altaffilmark{17},
F.~Menanteau\altaffilmark{14,15},
R.~Miquel\altaffilmark{28,16},
R.~L.~C.~Ogando\altaffilmark{12,13},
K.~Reil\altaffilmark{11},
A.~Roodman\altaffilmark{10,11},
E.~S.~Rykoff\altaffilmark{10,11},
E.~Sanchez\altaffilmark{29},
V.~Scarpine\altaffilmark{5},
R.~Schindler\altaffilmark{11},
M.~Schubnell\altaffilmark{23},
I.~Sevilla-Noarbe\altaffilmark{29},
M.~Smith\altaffilmark{30},
R.~C.~Smith\altaffilmark{4},
M.~Soares-Santos\altaffilmark{5},
F.~Sobreira\altaffilmark{12,31},
E.~Suchyta\altaffilmark{32},
M.~E.~C.~Swanson\altaffilmark{15},
G.~Tarle\altaffilmark{23}
\\ \vspace{0.2cm} (DES Collaboration) \\
}

\altaffiltext{1}{Department of Physics and Astronomy, University of Pennsylvania, Philadelphia, PA 19104, USA}
\altaffiltext{2}{Department of Astrophysical Sciences, Princeton University, Peyton Hall, Princeton, NJ 08544, USA}
\altaffiltext{3}{Jet Propulsion Laboratory, California Institute of Technology, 4800 Oak Grove Dr., Pasadena, CA 91109, USA}
\altaffiltext{4}{Cerro Tololo Inter-American Observatory, National Optical Astronomy Observatory, Casilla 603, La Serena, Chile}
\altaffiltext{5}{Fermi National Accelerator Laboratory, P. O. Box 500, Batavia, IL 60510, USA}
\altaffiltext{6}{LSST, 933 North Cherry Avenue, Tucson, AZ 85721, USA}
\altaffiltext{7}{CNRS, UMR 7095, Institut d'Astrophysique de Paris, F-75014, Paris, France}
\altaffiltext{8}{Department of Physics \& Astronomy, University College London, Gower Street, London, WC1E 6BT, UK}
\altaffiltext{9}{Sorbonne Universit\'es, UPMC Univ Paris 06, UMR 7095, Institut d'Astrophysique de Paris, F-75014, Paris, France}
\altaffiltext{10}{Kavli Institute for Particle Astrophysics \& Cosmology, P. O. Box 2450, Stanford University, Stanford, CA 94305, USA}
\altaffiltext{11}{SLAC National Accelerator Laboratory, Menlo Park, CA 94025, USA}
\altaffiltext{12}{Laborat\'orio Interinstitucional de e-Astronomia - LIneA, Rua Gal. Jos\'e Cristino 77, Rio de Janeiro, RJ - 20921-400, Brazil}
\altaffiltext{13}{Observat\'orio Nacional, Rua Gal. Jos\'e Cristino 77, Rio de Janeiro, RJ - 20921-400, Brazil}
\altaffiltext{14}{Department of Astronomy, University of Illinois, 1002 W. Green Street, Urbana, IL 61801, USA}
\altaffiltext{15}{National Center for Supercomputing Applications, 1205 West Clark St., Urbana, IL 61801, USA}
\altaffiltext{16}{Institut de F\'{\i}sica d'Altes Energies (IFAE), The Barcelona Institute of Science and Technology, Campus UAB, 08193 Bellaterra (Barcelona) Spain}
\altaffiltext{17}{George P. and Cynthia Woods Mitchell Institute for Fundamental Physics and Astronomy, and Department of Physics and Astronomy, Texas A\&M University, College Station, TX 77843,  USA}
\altaffiltext{18}{Department of Physics, IIT Hyderabad, Kandi, Telangana 502285, India}
\altaffiltext{19}{Institut de Ci\`encies de l'Espai, IEEC-CSIC, Campus UAB, Carrer de Can Magrans, s/n,  08193 Bellaterra, Barcelona, Spain}
\altaffiltext{20}{Kavli Institute for Cosmological Physics, University of Chicago, Chicago, IL 60637, USA}
\altaffiltext{21}{Instituto de Fisica Teorica UAM/CSIC, Universidad Autonoma de Madrid, 28049 Madrid, Spain}
\altaffiltext{22}{Department of Astronomy, University of Michigan, Ann Arbor, MI 48109, USA}
\altaffiltext{23}{Department of Physics, University of Michigan, Ann Arbor, MI 48109, USA}
\altaffiltext{24}{Center for Cosmology and Astro-Particle Physics, The Ohio State University, Columbus, OH 43210, USA}
\altaffiltext{25}{Department of Physics, The Ohio State University, Columbus, OH 43210, USA}
\altaffiltext{26}{Astronomy Department, University of Washington, Box 351580, Seattle, WA 98195, USA}
\altaffiltext{27}{Australian Astronomical Observatory, North Ryde, NSW 2113, Australia}
\altaffiltext{28}{Instituci\'o Catalana de Recerca i Estudis Avan\c{c}ats, E-08010 Barcelona, Spain}
\altaffiltext{29}{Centro de Investigaciones Energ\'eticas, Medioambientales y Tecnol\'ogicas (CIEMAT), Madrid, Spain}
\altaffiltext{30}{School of Physics and Astronomy, University of Southampton,  Southampton, SO17 1BJ, UK}
\altaffiltext{31}{Universidade Federal do ABC, Centro de Ci\^encias Naturais e Humanas, Av. dos Estados, 5001, Santo Andr\'e, SP, Brazil, 09210-580}
\altaffiltext{32}{Computer Science and Mathematics Division, Oak Ridge National Laboratory, Oak Ridge, TN 37831}
 

\begin{abstract}
We characterize the ability of the Dark
Energy Camera (DECam) to perform relative astrometry across its
500~Mpix, 3-deg$^2$ science field of 
view, and across 4 years of operation.  This is done using
internal comparisons of $\sim4\times10^7$ measurements of high-$S/N$
stellar images obtained in repeat visits to fields \edit{of moderate stellar
density,} with the
telescope dithered to move the sources around the array.   An empirical
astrometric model includes terms for: optical distortions; stray electric fields in the
CCD detectors; chromatic terms in the instrumental and atmospheric
optics; shifts in CCD relative positions of up to $\approx10\,\mu$m
when the DECam temperature cycles; and
low-order distortions to \edit{each} exposure from changes in atmospheric refraction and
telescope alignment. 
Errors in this astrometric model
are dominated by stochastic variations with typical amplitudes
of 10--30~mas (in a 30 s exposure) and 5\arcmin--10\arcmin\ coherence
length, plausibly attributed to Kolmogorov-spectrum atmospheric
turbulence.  The size of these atmospheric distortions is not closely
related to the seeing.
Given an astrometric reference catalog at density
$\approx0.7~\textrm{arcmin}^{-2},$ \eg\ from Gaia, the typical atmospheric
distortions can be interpolated to $\approx7$~mas RMS accuracy (for
30~s exposures) with
$1\arcmin$ coherence length \edit{in for residual errors}.
Remaining detectable error contributors are 2--4~mas RMS from unmodelled
stray electric fields in the devices, and another 2--4~mas RMS from focal
plane shifts between camera thermal cycles.  Thus the astrometric
solution for a single \edit{DECam exposure is accurate to
3--6~mas ($\approx0.02$~pixels, or $\approx300$~nm) on the focal
plane, plus the stochastic atmospheric distortion.}
\end{abstract}


\section*{}
\renewcommand{\baselinestretch}{0.75}\normalsize
\tableofcontents
\renewcommand{\baselinestretch}{1.0}\normalsize
\newpage

\section{Introduction}
The Dark Energy Camera \citep[DECam]{decam} is one member of a new generation
of high-throughput imagers combining large field of view (3 deg$^2$ for
DECam) with large telescope aperture (the 4-meter Blanco telescope).
In the post-Gaia era \citep{gaia}, when positions, proper motions, and parallaxes
are expected to be available with $<1$~milliarcsecond (mas) accuracy 
for the $10^9$ stars with magnitude $G\lesssim 20$,  what need do we
have for accurate astrometry from these large ground-based cameras?
There remain strong scientific motivations to obtain the most accurate
possible positions for sources fainter than Gaia's limit, for
transient sources, and for solar-system bodies.  Ideally these
general-purpose, large-format imagers would be capable of obtaining
astrometric measurements limited by the unavoidable shot noise and
atmospheric fluctuations.  Motivation and practice of astrometry from
large-format ground-based CCD cameras have been discussed by
\citet{anderson}, \citet{platais}, \citet{bouy}, and \citet{magnier},
among others.  Accurate astrometry underlies
many of the science goals of the \textit{Large Synoptic Survey
  Telescope (LSST)} now under construction \citep{lsstbook}.

In addition, one of
the motivators for construction of DECam is measurement of weak
gravitational lensing distortions of galaxies.  Success in this
pursuit requires the ability to register multiple exposures of every
galaxy to an accuracy of $\approx10$~mas or better---otherwise the
blur induced by misregistration in combining images could be 
mistaken for a coherent weak-lensing distortion.  Searches for
transient sources also benefit from precise image registration to
improve subtraction of static sources.


One thing we do \emph{not} need our wide-field imagers to do is
determine absolute positions, since the preliminary Gaia catalogs
are sufficiently dense to yield thousands of stars in the spatial and
dynamic-range overlap between Gaia and most DECam exposures.  These
suffice to determine the absolute pointing and any low-order
astrometric distortion terms across the DECam field of view.  In this
work we therefore focus on establishing \emph{relative} astrometry
with DECam on scales $\lesssim1\arcdeg$.  
Indeed one might ask why to bother at all with the effort making an
astrometric model for DECam instead of simply interpolating all
positional errors from Gaia stars.  First, many
of the detector-level effects occur on angular scales too small for Gaia stars
to sample.  Second, if our model removes discontinuities in the
astrometric errors between CCDs, we can interpolate using reference
stars from the whole field rather than being confined to those on a
single device.  Furthermore the Gaia proper motion catalog is not
yet available, so the reference catalog is not yet at mas accuracy.
Lastly many DECam exposures may have dynamic range which does not overlap
well with the Gaia catalog.

In Section~\ref{methods} we
describe our method of deriving the DECam astrometric map and its
error properties by forcing internal agreement amongst stellar
positions in a series of offset exposures of rich star fields.  In
Sections~\ref{data}, \ref{model}, and \ref{residsec} we describe the 
data used to characterize DECam astrometry, the model applied to it,
and the \emph{static} residuals  to this model,
\ie\ those which repeat from
exposure to exposure.  Section~\ref{atmosphere} characterizes the
\emph{stochastic} residuals, \ie\ those uncorrelated between
consecutive exposures and presumably due to atmospheric fluctuations.
Section~\ref{stability} characterizes the changes in the astrometric
model from night to night and over the first 4 years of DECam
observations.  Section~\ref{interpolation} investigates how much of the
stochastic distortion can be removed by interpolation from a set of
reference stars of a given density.

Our goal will be to model any astrometric distortion that contributes
more than $\approx 1$~mas RMS error that is correlated between stars separated
by $>10\arcsec$.  To put this scale in context, note that the mean
scale of a 15~$\mu$m  DECam pixel is 264~mas, so 1~mas corresponds to
0.004 pixel $=60$~nm, or about 100 atoms in the silicon lattice.
The DECam science array consists of 62
deep-depletion CCDs, each $2048\times4096$ pixels, and the array spans
a roughly hexagonal area of diameter 2\arcdeg.  Thus 1~mas is 1.4
parts in $10^7$ of the DECam field of view.  Scale changes due to
stellar aberration, air
pressure variations, and atmospheric refraction are far larger than
this, so we will clearly need to allow each exposure an independent
overall linear transformation across the FOV to approach mas
accuracy.  Indeed the \emph{nonlinear} portion of atmospheric
refraction is expected to have peak-to-peak amplitude of $(14\,{\rm mas})\times \sec^2 z
\tan z$ across the DECam FOV (where $z$ is the zenith angle), so
we must allow at least quadratic freedom to our solution atop any
static instrumental model.

RMS positional errors reported in this paper refer to the sum of 
E-W and N-S components, unless noted otherwise.

\section{Methods}
\label{methods}
The astrometric solution for DECam is a parametric model
for the celestial (world) coordinates $\vx^w=(x^w,y^w)$ of
an object given its observed pixel coordinates $\vx^p=(x^p,y^p)$
and some set of observing circumstances $C$, which might include the
object's color $c$, plus discrete variables such as the date,
exposure, filter, and individual CCD on which the measurement was
obtained.  The solution is found by straightforward
$\chi^2$-minimization over the values of the model parameters
$\vpi.$  The index $i$ ranges over all position measurements used to
constrain the solution, and we assume a measurement error $\sigma^p_i$
that is the same for both positional components of $\vx^p_i$.  We
index the distinct objects on the sky by $\alpha$, let $\alpha_i$ be
the object targeted by measurement $i$, and denote by $i \in \alpha$
the subset of measurements with $\alpha_i=\alpha$.  We define
\begin{align}
\label{chisq}
\chi^2  & \equiv \sum_i
w_i\left| \vx^w(\vx^p_i, C_i, \vpi) - \bar \vx_{\alpha_i} \right|^2 \\
\bar \vx_{\alpha} & \equiv \frac{\sum_{i \in \alpha} w_i
  \vx^w(\vx^p_i, C_i, \vpi)}{\sum_{i \in \alpha} w_i} \\
\label{weights}
w_i^{-1} & = \left( \sigma_i^2 + \sigma_{\rm sys}^2\right) 
\left| \frac{d\vx^w}{d\vx^p}\right|_i.
\end{align}
In (\ref{weights}) we introduce $\sigma_{\rm sys}$ to prevent very
high weights from being assigned to high-$S/N$ measurements.  We may
consider $\sigma_{\rm sys}$ to represent the expected stochastic
position errors beyond those arising from image noise that are
included in $\sigma_i.$ In practice we find that stochastic
atmospheric distortions dominate the astrometric residuals, so we set
$\sigma_{\rm sys}$ near the typical RMS atmospheric distortion in our data
($\sim 10$~mas, \textit{cf.} Sec \ref{atmosphere}).  This fairly arbitrary choice appropriately equalizes the weights
assigned to individual measurements, but it does mean that our final
$\chi^2$ values should not be expected to follow a $\chi^2$
distribution---it serves only as the quantity used to optimize \vpi.

Note that measurements from DECam can be freely mixed with other instruments'
position measurements in \eqq{chisq}.
\emph{Internal} constraints---that multiple DECam observations of a
source yield the same \edit{world coordinates}---are combined with \emph{external}
constraints that DECam match a source of \textit{a priori} assigned
\edit{world coordinates} of these objects. We denote as a \textbf{reference
  catalog} any set of measured positions that are independent of \vpi,
\ie\ the \edit{ function $\vx^w(\vx^p)$ is simply the identity, and
  the reference catalog directly specifies $\vx^w_i$ and errors $\sigma^w_i.$}

Our strategy for DECam calibration is
to produce very strong internal constraints by taking a series of
$\approx20$ consecutive exposures of fields at modest Galactic
latitudes, where stellar sources are abundant but not crowded.  The
pointings of these exposures are shifted by anywhere from 10\arcsec\
to the FOV diameter, so that a given star is imaged at many places on
the array.  In this scheme the reference catalog serves mainly to
break degeneracies in overall position and linear scaling of the
astrometric map (see Section~\ref{degeneracies}).  These sequences of
exposures are called \textbf{star flats} (since they are also used to
calibrate photometric response).  Since DECam was installed in 2012,
star flat sequences in all filters have been executed several times
per year, usually during bright time.  These data, described in
Table~\ref{starflats}, are the ones used in this paper to derive the
DECam astrometric model. 

In the remainder of this section we will detail the algorithmic and
coding choices made in defining the maps $\vx^w(\vx^p)$ and in the
minimization of $\chi^2.$  \edit{These choices are realized in 
\texttt{C++} code with an executable program called \wcsfit.}
A reader uninterested in the implementation
details can skip to Section~\ref{model}.

\subsection{Terminology}
\label{terminology}
We adopt the following terminology:
\begin{itemize}
\item A {\bf pixel map} is a function $\vx^w(\vx^p)$ giving the
  \edit{transformation} from detector coordinates to world
  coordinates. \edit{These maps are realized by compounding several
    functions, each of which may also be referred to as a pixel map.}
\item A {\bf detection} is a single measurement of a stellar position,
  which as noted is described by pixel coordinates $\vx^p_i$ and an
  associated uncertainty $\sigma_i.$
\item A {\bf device} is a region of the focal plane over which we
  expect to have a continuous pixel map, \ie\ one of the CCDs
  in the DECam focal plane.  Every detection belongs to exactly one device.
\item An {\bf exposure} comprises all the detections obtained
  simultaneously during one opening of the shutter.  The exposure
  number is essentially our discrete time variable.
\item An {\bf extension} comprises the detections made on a single
  device in a single exposure.\footnote{The name arises from each
    device's detection list typically appearing in a distinct
    binary table extension of a FITS-format file.}  \wcsfit\ allows each extension to be
  assigned its own pixel map, which will be a continuous function.
  Every detection belongs to exactly one extension.
\item A {\bf catalog} is the collection of all detections from a
  single exposure, \ie\ the union of the extensions from all the
  devices in use for that exposure.
\item A {\bf band} labels the filter used in the observation.  Every
  exposure has exactly one band.
\item An {\bf epoch} labels a range of dates over which the physical
  configuration of the instrument, aside from filter choice and the
  pointing of the telescope, is considered (astrometrically)
  invariant.  Every exposure belongs to exactly one epoch.
\item An {\bf instrument} is a given configuration of the camera for which we expect
  the instrumental optics to yield an invariant astrometric solution.
  In our analyses an instrument is specified by a combination of band
  and epoch.  Every exposure is associated with exactly one instrument.
  \edit{This is the same definition as used in \scamp\ 
    \citep{scamp}, the public code commonly used for astrometric solutions.}
\item A {\bf field} is a region of the sky holding the detections from
  a collection of exposures.  Every exposure is associated with exactly
  one field.
Each field $f$ has a central right
  ascension and declination $(\alpha_f, \delta_f)$.  The world
  coordinates $\vx^w$ are defined to be in the gnomonic projection of
  the sky about the field center.
\item A {\bf match}, sometimes called an {\bf object}, comprises all
  the detections that correspond to a common celestial source and are
  therefore expected to have common true $\vx^w.$\footnote{\wcsfit\
    does not yet consider proper motions of sources.}  In
  this astrometric study we will make use only of stellar sources, so
  a match is simply a star.  We only allow matches to be constructed
  between detections in a common field.
\item A {\bf reference catalog} is an extension for which there are no
  free parameters in the map to \edit{world coordinates,} for example the list of
  Gaia stars for a given field.  The distinction between devices,
  instruments, etc. is irrelevant for these, and we can consider all sources of
  reference information as belonging to a common catalog.
\end{itemize}

As detailed in Section~\ref{maps}, \wcsfit\ allows the pixel map for
extension $k$ to be composed of a sequence of ``atomic''
transformations.  
Following \scamp\ we will divide the \edit{overall pixel
  map} applied to a
given extension into an \emph{instrumental} map
followed by an \emph{exposure} map.  The former goes from pixel
coordinates of each device to an intermediate system of a gnomonic
projection about the telescope optic axis, and is taken to be constant
within an epoch.  The exposure map is continuous across the field of
view and takes independent parameters for each exposure.

\subsection{Available maps}
\label{maps}
We must specify a functional form (and free parameters) for a map
$\vm_k(\vx^p,c)$ from pixels to \edit{world} coordinates for each extension
$k$.  Here $c$ is the object color, and the other elements of the
observational circumstances $C$ are specified by the extension
index. \wcsfit\ allows each map $\vm_k$ to be specified as the composition of
a series of $j=1,2,\ldots,N_k$ ``atomic'' coordinate transformations $\vm^{(i)}_k$:
\begin{align}
\label{compose1}
\vx^{(0)} & = \vx^p, \\
\label{compose2}
\vx^{(i)} & = \vm^{(i)}_k(\vx^{(i-1)}) \\
\label{compose3}
\vx^w & = \vx^{(N_k)}.
\end{align}
We will generically refer to the input of each transformation
$\vm^{(i)}_k$ as its ``pixel'' coordinates and the output as its
``world'' coordinates, even though the intermediate variables are in
fact neither. \edit{In our application, the chain of component maps is
  divided into those of the instrument solution followed by those of
  the exposure solution.}

\edit{\wcsfit\ follows the definitions in Section~\ref{terminology} by making
each coordinate transformation or element thereof an}
instance of an abstract \texttt{C++} base class \texttt{PixelMap}.
Each has a \texttt{type}, a unique \texttt{name} string, and has a
number $\ge0$ of free parameters controlling its
actions.  \texttt{PixelMap} instances can be (de-)serialized (from) to
ASCII files in \textsc{YAML} format, easily read or written by
humans.  The \wcsfit\ user specifies the transformations to be fit to
the data by giving the program such a \textsc{YAML} file as
input---the parameters are assigned default starting values if none
are specified.  Anywhere that the strings
\texttt{BAND,INSTRUMENT,EPOCH,} or \texttt{DEVICE} appear in these
input files they are replaced with the values appropriate to the
extension, allowing a generic model to be specified compactly.  The
\wcsfit\ user can also specify the names of any \texttt{PixelMaps}
whose parameters should be held fixed at their input values.
The primary output of \wcsfit\ is another \textsc{YAML} file
specifying all of the maps and their best-fit parameters.

The types of \texttt{PixelMaps} available for use are:
\begin{itemize}
\item The \texttt{Identity} map, which leaves $\vx$ unchanged, and has
  no free parameters.
\item \texttt{Constant} maps have $\vx^w = \vx^p + \vx_0,$ with the
  two components of $\vx_0$ as parameters.
\item \texttt{Linear} maps have $\vx^w = A\vx^p + \vx_0,$ with six
  parameters in $\vx_0$ and the components of the matrix $A$.
\item \texttt{Polynomial} maps have their free parameters as the
  coefficients of two polynomials of specified degrees $d_x$ and $d_y$
  in the $\vx^p$ components that produce $x^w$ and $y^w$, respectively.
\item \texttt{Template} maps apply transformations based on lookup
  tables.  One has the option of x, y, or radial transformations:
\begin{align}
  x^w & = x^p + s f(x^p), \\
  y^w & = y^p + s f(y^p), \text{or} \\
  \vx^w & = \vx^p + s \frac{\vx^p-\vx_c}{|\vx^p-\vx_c|} f\left(
          |\vx^p-\vx_c| \right),
\end{align}
\edit{The first two cases each operate in only a single cartesian
  direction.  In the third (radial)}
case, the center $\vx_c$ of the distortion is
specified.  There is a single free parameter, the scaling parameter
$s$.  The template function $f$ is defined as linear interpolation
between values $v_j$ at nodes $a_0 + j\,\Delta a$ for $0\le j \le N$.
\item \texttt{Piecewise} maps are functionally identical to the
  \texttt{Template} map, except that the nodal values $v_j$ are the
  free parameters, and the scaling is fixed to $s=1.$  
\item A \texttt{Color} term is defined by 
\begin{equation}
\vx^w = \vx^p + \left(c-c_{\rm
    ref}\right)\left[\vm(\vx^p)-\vx^p\right]
\end{equation}
where $c_{\rm ref}$ is a reference color and $\vm$ is an instance of
any of the above forms of transformation.  The parameters of the
\texttt{Color} map are those of the map it scales.
\item \texttt{Reprojection} maps have no free parameters: they merely
  move coordinates from one projection of the sphere to another.  
\item \texttt{Composite} maps realize
  Equations~(\ref{compose1})--(\ref{compose3}) for a specified
  sequence of any of maps (including other \texttt{Composite} maps).
  The parameters of the composite are the concatenation of those of
  the component maps.
\end{itemize}

A \texttt{PixelMap}, in combination with a specification of the
projection in which the $\vx^w$ maps to the celestial sphere, forms a
complete world coordinate system (WCS).

\subsection{Degeneracies}
\label{degeneracies}
When minimizing $\chi^2$ we must be aware of degeneracies whereby
\vpi\ can change while $\chi^2$ is invariant.  Such degeneracies will
lead to (near-)zero singular values in the normal matrix \matA\ used
in the solution for \vpi\ (Section~\ref{model}), and failures or
inaccuracies in its inversion.  There are several such landmines which
we must clearly avoid.  We will assume in this discussion that the
astrometric model for each extension is a
device-based instrumental function $\vx^T=D(\vx^p)$ from pixel to
``telescope'' coordinates, followed with an exposure-based function
$\vx^w = E(\vx^T).$

\subsubsection{Shift}
The simplest degeneracy is a shift in all stellar positions,
$E\rightarrow E+\Delta\vx$ for every exposure (in the flat-sky limit;
more generally the degeneracy is a rotation of the celestial sphere).
Each star $\alpha$ has its derived sky position $\bar\vx^w_\alpha$
shifted as well, but since $\chi^2$ is differential, there is no
effect on $\chi^2.$  This degeneracy is broken by having a reference
catalog for which $\vx^w$ is fixed.  The reference catalogs does not
need to be very dense or precise to break this degeneracy.

\subsubsection{Color shift}
A color-dependent shift $E\rightarrow E + c\, \Delta\vx$ is also
undetectable in the differential $\chi^2$.  This degeneracy is broken
if colors are known for reference stars over a finite range of color.

\subsubsection{Linear}
In the flat-sky limit consider the case where the exposure component
for exposure $k$ is an affine transformation $E_k = \matA_k \vx^T +
\vx_k$ with linear rescaling $\matA_k$ and offset $\vx_k$, the latter
corresponding to the pointing of the telescope at exposure $k$.  An
object $\alpha$ with world coordinates $\vx^w_\alpha$ will be observed
at telescope coordinate $\vx^T_{k\alpha} =
\matA^{-1}\left(\vx^w_\alpha - \vx_k\right).$  For any non-degenerate
matrix $\matB$ there is an alternative solution
\begin{align}
\matA_k & \rightarrow \matB \matA_k \\
\vx_k & \rightarrow \matB^{-1} \vx_k \\
\vx^w_\alpha & \rightarrow \matB \vx^w_\alpha
\end{align}
which leaves $\chi^2$ unchanged.  This degeneracy is also broken by
the existence of a sample of reference \edit{stars}.  There is also a
color-dependent variant of this degeneracy.

If the exposure solution $E_k$ has freedom to be altered by some global
polynomial function $B$ of order $n$, then there is generalization of
this degeneracy in which each $E_k$ is shifted by a polynomial of
order $n-1$.  Again the solution is to have a reference catalog of
even modest density and accuracy.

\subsubsection{Colony collapse disorder}
\wcsfit\ is accelerated by calculating the weight of each observation
in \eqq{weights} just once at the start of fitting, using the
determinant of the starting WCS system to convert the pixel errors
into world coordinate errors.  This opens the door to a
pseudo-degeneracy in which all output $\vx^w$ values are scaled by
some matrix $\matB,$ sending $\chi^2\rightarrow |\matB| \chi^2.$ If
$|B|\rightarrow 0$, the solution appears to approach perfection while
collapsing the output map.
This is
countered by an increase in $\chi^2$ contributed by the reference
stars, which are not collapsing; but if the total weight of the reference stars is too low, the
solution will tend toward collapse.  The collapse becomes complete if
the reference stars are then flagged as outliers and removed by our
$\sigma$-clipping step.  \wcsfit\ includes a parameter to scale the
weights of the reference catalogs, which can be used to prevent this
collapse solution if the reference catalog is sparse.

\subsubsection{Exposure/instrument trades}
For any map $F$, the transformations
\begin{align}
E_k & \rightarrow E_k F \\
D & \rightarrow F^{-1} D
\end{align}
clearly leaves $\chi^2$ and all $\vx^w$ values invariant.  If the
functional forms being used for $E$ and $D$ admit such a
transformation, then the solution is degenerate.  The \wcsfit\ code
searches for cases where multiple \texttt{Constant, Linear,} or
\texttt{Polynomial} atomic map elements are composited into any
exposures' pixel maps and are hence able to trade their terms.
This degeneracy can be broken by setting one of the exposure maps
$E_k$ to the \texttt{Identity} map.  \wcsfit\ will do this
automatically if the user's configuration leaves such degeneracies in place.

\subsubsection{Unconstrained parameters}
Map parameters are of course degenerate if there are no stellar
observations being affected by them---{\it e.g.} if a given exposure
did not generate any matched detections, or they have all been removed
as outliers, then the parameters of the exposure solution are
unconstrained.

\wcsfit\ checks the normal matrix \matA\ for null rows that arise
when a parameter does not act on any observations.  In this case the
diagonal element on this row is set to unity, which stabilizes the
matrix inversion and freezes this (irrelevant) parameter in further
iterations.

More troublesome is the case where there are a small but non-zero number
of observations on an exposure, too few to constrain the model, so
that \matA\ is degenerate but without null rows.  In this case
\wcsfit\ will fail the attempt to do a Cholesky decomposition of the
non-positive-definite  \matA.  In this case \wcsfit\ will perform a
singular value decomposition of \matA, report to the user which
parameters are associated with near-null singular values, \edit{then quit.}

\subsection{Algorithms}

The \wcsfit\ software suite assumes that we are already in possession
of an initial WCS for each extension of sufficient accuracy to allow
unambiguous matching of common detections of a source.  \scamp\ is
routinely run on each DES exposure to generate this starting WCS, with
accuracy of $< 1\arcsec$ relative to Gaia or other reference
catalog.

\subsubsection{$\chi^2$ minimization}
Another benefit of having a good starting WCS for each exposure is
that we can initialize parameters of the maps that are defaulted on
input by fitting them to the starting WCS---\wcsfit\ generates set of
pseudo-detections on a grid of $\vx^p$ spanning the device, and fits
them to a pseudo-reference catalog holding the $\vx^w$ positions to
which the pixel positions are mapped by the WCS.  

The algorithm for 
minimization of $\chi^2$ assumes that the minimizing solution is close
to the starting solution, \ie\ we are doing fine tuning after \scamp\
has done the work of bringing us close.  The positions are close to
linear in the parameters, so the $\chi^2$ value should be close to the
usual quadratic form
\begin{align}
\chi^2 & \approx \chi^2(\vpi_0) + 2{\bf b} \cdot \Delta\vpi +
         \Delta\vpi \cdot \matA \cdot \Delta \vpi, \\
 b_\mu & \equiv \frac{1}{2} \frac{\partial \chi^2}{\partial \pi_\mu} = \sum_i 
         w_i\left(\vx^w(\vx^p_i, \vpi_0) - \bar \vx_{\alpha_i} \right)
  \cdot \left(\frac{\partial \vx^w(\vx^p_i, \vpi)}{\partial \pi_\mu}
   - \frac{\partial \bar \vx_{\alpha_i}}{\partial \pi_\mu} \right)\\
A_{\mu\nu} & \equiv \left(\frac{\partial \vx^w(\vx^p_i, \vpi)}{\partial \pi_\mu}
   - \frac{\partial \bar \vx_{\alpha_i}}{\partial \pi_\mu} \right)
             \cdot 
\left(\frac{\partial \vx^w(\vx^p_i, \vpi)}{\partial \pi_\nu}
   - \frac{\partial \bar \vx_{\alpha_i}}{\partial \pi_\nu} \right).
\end{align}
Note that the weights $w_i$ are being assumed independent of \vpi,
\ie\ the world-coordinate errors $\sigma^w_i$ of each exposure are
held fixed at the values implied by the starting WCS.  Also note that
\wcsfit\ does not treat the true positions $\vx_\alpha^w$ of the
sources as free parameters.  Instead the dependence of the mean of the
measurements $\bar \vx_\alpha$ upon the parameters is propagated
directly into the normal equation.  

The calculation of ${\bf b}$ and \matA\ is the most computationally
intensive part of \wcsfit.  The summation for matches is distributed
across cores using \textsc{OpenMP} calls.  Each match is dependent
upon the limited subset of the parameters \vpi\ which appear in the
pixel maps for the extensions in which the object is observed, hence
the updates to \matA\ are sparse, though the final matrix is dense.

\wcsfit\ first attempts the Newton iteration
\begin{equation}
\label{Newton}
\vpi \rightarrow \vpi - \matA^{-1} {\bf b}.
\end{equation}
The solution time scales as the cube of the number of free
parameters, and is executed using a multithreaded Cholesky
decomposition after preconditioning \matA\ to have unit diagonal
elements.  Despite the cubic scaling, this step is usually faster 
than the calculation of the normal matrix. If the decomposition fails due to a
non-positive-definite \matA, \wcsfit\ performs a singular-value
decomposition on \matA\ and informs the user which parameters dominate
the degenerate vectors.

The Newton step is iterated until $\chi^2$ no longer decreases by more
than a chosen fraction.  Should $\chi^2$ increase during an iteration,
or fail to converge within a selected number of steps, then the
minimization process is re-started using a Levenberg-Marquart
algorithm based on the implementation by \citet{recipes}.

\subsubsection{Outlier rejection}
The \wcsfit\ solutions must be robust to astrometric measurements
perturbed by unrecognized cosmic rays or defects on the stellar
images, and by stars with proper motion or binary partners which alter
the photocenter by amounts exceeding measurement errors.  We do not at
this time fit for proper motion or parallax within \wcsfit.

Outlier rejection is done using standard $\sigma$-clipping
algorithms. A clipping threshold $t$ is specified at input.  After
each $\chi^2$ minimization, a rejection threshold is set at $t
\sqrt{\chi^2/\textrm{DOF}}$.  Detections whose residual to the fit (in
units of $\sigma$) exceeds the threshold are discarded.  At most one
outlier per match is discarded at each clipping iteration.

Outlier clipping is alternated with $\chi^2$ minimization until the
clipping step no longer reduces the $\chi^2$ per degree of freedom by
a significant amount.

\subsubsection{Procedure}
The steps in the astrometric solution process are as follows:
\begin{enumerate}
\item A preparatory Python program 
reads an input YAML configuration file specifying the desired input
catalog files, 
plus the definitions of the fields, epochs, and instruments.  It then
collects from all the catalogs and their headers any information
necessary to construct tables of extensions, 
devices, exposures, and instruments.  This includes extracting the
serialized starting WCS, usually as produced by \scamp\ and stored in
the headers of the FITS catalog extensions.  
\item A second preparatory program reads all the detections from the
  input catalogs, applying any desired cuts for $S/N$ and stellarity,
  and then runs
a standard friends-of-friends algorithm to identify all matching
detections.  Any match that includes multiple
detections from the same exposure is discarded.  The id's of all
groups of matching detections are then stored in another FITS table.

\item \wcsfit\ starts by ingesting the input FITS tables and creating
  the structures defining instruments, devices, exposures, and
  extensions.
\item The YAML file specifying the pixel maps to be applied to each
  extension is parsed, and a \texttt{PixelMap} is created with
  specified or defaulted parameters.  Any of the map elements may have
  its parameters frozen by the user, the remainder are the free
  parameters of our model.
\item \wcsfit\ checks the map configuration for degeneracies: is there
  reference catalog in each field? Are there are any
  exposure/instrument degeneracies?  If so, \wcsfit\ will attempt to
  break the degeneracies by setting one or more exposures' maps to
  \texttt{Identity}. 
\item All exposures in a field are reprojected to a common gnomonic
  system about the field center.
\item Any parameters of \texttt{PixelMaps} that were 
  set to defaults have their values set by a least-squares fit to
  the starting WCS.  Any degeneracies halt the program.
\item The $\vx^p_i$ and $\sigma^p_i$ of all detections that are part
  of useful matches are extracted from their source catalogs.  For any
  detections whose maps include color terms, we require a measurement
  from a color catalog to be matched to the same object.  The color
  catalog is read at this point.
\item A requested fraction of the matches are excluded from the fit at
  random.  These reserved matches can be used later to validate the fit.
\item Any exposures containing insufficient detections are removed
  from the fit.
\item The iteration between $\chi^2$ minimization and
  $\sigma$-clipping begins.  At each iteration, \matA\ is checked for
  null rows as noted in Section~\ref{degeneracies}, which are altered
  so as to freeze the associated parameter.  If \matA\ is not
  positive-definite, \wcsfit\ reports the nature of the associated
  degenerate parameters, then exits.
\item The best-fit astrometric model is written to an output YAML file.
\item After completion of the fit, the best-fit map is applied to both
  the fit and reserved matches.  The $\sigma$-clipping algorithm is
  applied iteratively to the reserved matches.
\item The RMS residual and $\chi^2$ statistics are reported for the
  un-clipped detections on each exposure.
\item The input, output, and best-fit residual for every detection are
  written to an output FITS table for further offline analyses.
\end{enumerate}

\subsection{Performance}
The run of \wcsfit\ producing the results in Section~\ref{model} was
executed on a dual-CPU workstation with a total of 12 2.4~GHz cores.
After reserving 30\% of the matches, we fit 19~million detections in
311,000 distinct matches.  There are 4948 map elements with a total of
26,645 free parameters.  Each calculation of \matA\ takes
approximately one hour, and the linear solution takes one minute.
Five iterations of minimization/clipping were required for
convergence.

\section{Input data}
\label{data}
\begin{figure}[ht]
\center
\includegraphics[width=0.5\columnwidth]{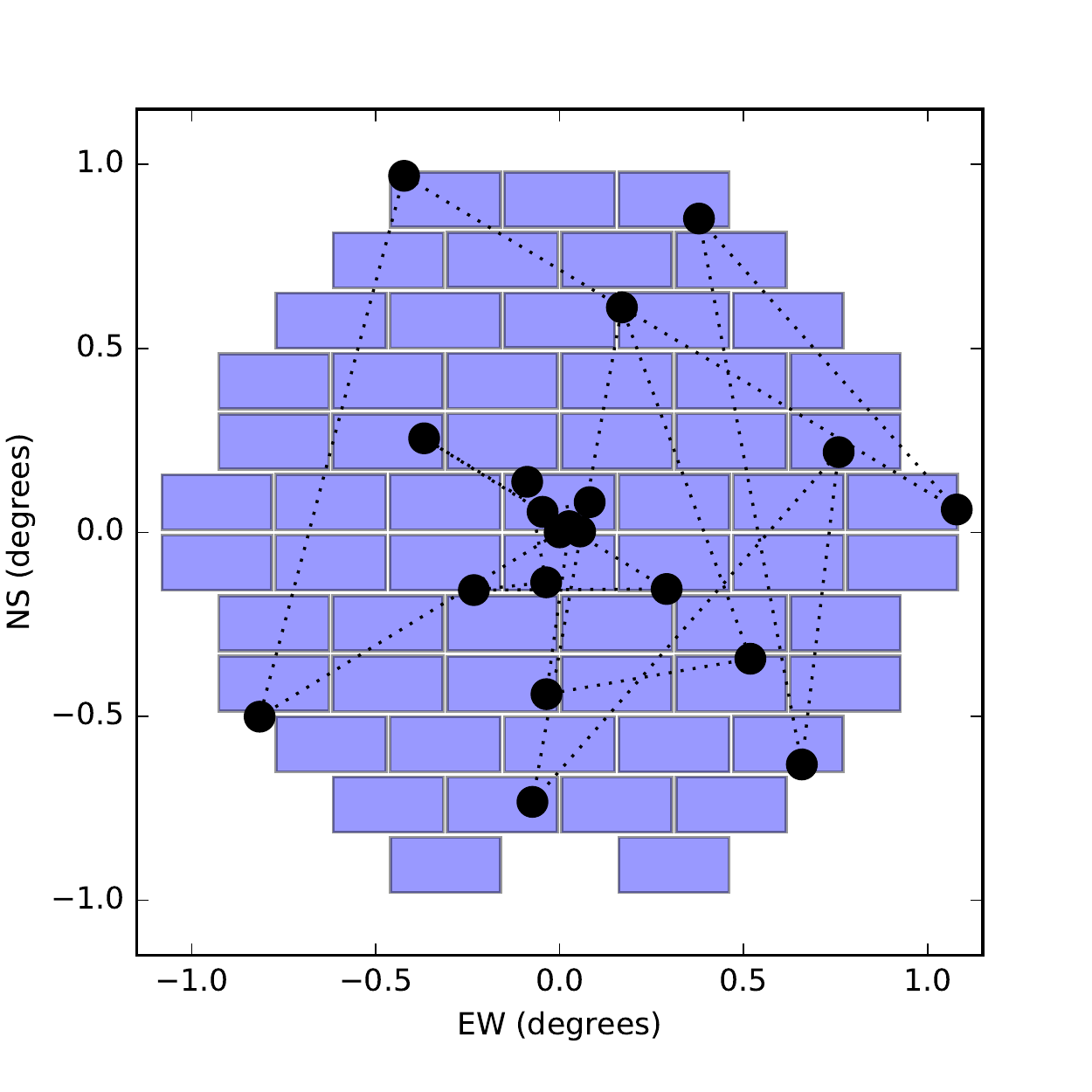}
\caption[]{\small The dots show the pointing positions for a typical
  series of exposures in a single filter for a star flat sequence.
  These are overlain on an outline of the 61 functional DECam science
  CCDs as of December 2012.  The dashed line connects the pointings in
  the order they are exposed.}
\label{dithers}
\end{figure}

The astrometric solution is derived from multiple epochs of the star
flat observations described above.  Table~\ref{epochs} lists the dates
and conditions of the star flat sequences during the first four years of DECam
operations for which there were neither
clouds nor instrument anomalies.\footnote{Note that the number of
  functional CDDs on DECam dropped from 61 to 60 after one year of
  operation.  Plots in this paper hence vary in the number of CCDs in use.}
  functional CCDs at the start of   Exposures are usually 30~s long,
with 25--30~s dead time 
for readout and repointing, so the star flat sequence for 5 filters
consumes about 2~hours of clock time.  Figure~\ref{dithers} shows a
typical star flat pointing sequence of 22 exposures.  \edit{There are $\approx10^5$
  Gaia reference stars available in the area covered by the star flat
  exposures of a given field.}

Each exposure is run through the standard DES data reduction pipeline,
including linearization of images, crosstalk removal, correction for
the ``brighter-fatter effect'' \citep{gruen}, debiasing, and
division by dome flats, and subtraction of sky and fringe signals.
Sources are detected and measured using \textsc{SExtractor}
\citep{sextractor}.  For the following analyses we filter the catalogs
for sources with no \textsc{SExtractor} flags set, no defective,
saturated, or cosmic-ray-flagged pixels
within the isophote, with
\texttt{MAGERR\_AUTO}$<0.01,$ indicating signal-to-noise ratio
$S/N\gtrsim100,$ and with $|\texttt{SPREAD\_MODEL}|<0.003$ to select
only stellar sources.  The flag cut removes objects that overlap
detected neighbors.  

The windowed centroids
$(\texttt{XWIN\_IMAGE},\texttt{YWIN\_IMAGE})$ are used for centroid
positions,  $\vx^p$, as they have been demonstrated to be robust to
the details of the point-spread function (PSF) while approaching the
accuracy of ideal PSF-fitting astrometry. Our focus on astrometric errors that
correlate over space and/or time means we will not
investigate the vagaries of centroid measurement, \eg\ pixel-phase errors.

The density of useful stellar positions varies with field,
filter, seeing, and sky conditions, but is usually 200-400 per CCD, or
more than 10,000 per exposure and $>10^6$ per star flat epoch.

\begin{deluxetable}{lccc}
\tabletypesize{\footnotesize}
\tablewidth{0pt}
\tablecolumns{8}
\tablecaption{Star flat observing sequences and DECam thermal events
  through Sep 2016\label{epochs}}
\tablehead{
\colhead{Epoch\tablenotemark{a}} & 
\colhead{Field} &
\colhead{$D_{50}$\tablenotemark{b}} &
\colhead{Airmass}
}
\startdata
\texttt{20121120}\tablenotemark{c} &  0640--3400 & 2\farcs09 & 1.04 \\
\texttt{20121223} &   0730--5000 & 1\farcs04 & 1.06 \\[3pt]
\textit{2012 Dec 30} & \multicolumn{3}{c}{\textit{Camera warmup}} \\[3pt]
\texttt{20130221} &   1327--4845 & 1\farcs12 & 1.06 \\[3pt]
\textit{2013 May 12}  & \multicolumn{3}{c}{\textit{Camera warmup}} \\
\textit{2013 July 22} & \multicolumn{3}{c}{\textit{Camera warmup}} \\[3pt]
\texttt{20130829} &   1900--5000 & 1\farcs10 & 1.07 \\[3pt]
\textit{2013 Oct 15} & \multicolumn{3}{c}{\textit{Camera warmup}} \\[3pt]
\texttt{20131115} &   0640--3400& 1\farcs41 & 1.09 \\[3pt]
\textit{2013 Nov 30} & \multicolumn{3}{c}{\textit{CCD S30 fails}} \\[3pt]
\texttt{20140118} &   1327--4845 & 1\farcs33 & 1.33 \\[3pt]
\textit{2014 May 12}  & \multicolumn{3}{c}{\textit{Camera warmup}} \\[3pt]
\texttt{20140807}\tablenotemark{d} &   1327--4845 & 1\farcs43 & 1.32 \\
\texttt{20141105} &   0640--3400 & 1\farcs28 & 1.01 \\[3pt]
\textit{2014 Dec 1}  & \multicolumn{3}{c}{\textit{Camera warmup}} \\[3pt]
\texttt{20150204} &   1327--4845 & 0\farcs88 & 1.31 \\[3pt]
\textit{2015 May 25}  & \multicolumn{3}{c}{\textit{Focal plane
    temperature drop}} \\
\textit{2015 June 25}  & \multicolumn{3}{c}{\textit{Partial camera warmup}} \\
\textit{2015 July 25}  & \multicolumn{3}{c}{\textit{Camera warmup}} \\
\textit{2015 Aug 9}  & \multicolumn{3}{c}{\textit{Camera warmup}} \\
\textit{2015 Aug 25}  & \multicolumn{3}{c}{\textit{Camera warmup}} \\[3pt]
\texttt{20150926} &   2040--3500 & 1\farcs19 & 1.01 \\[3pt]
\textit{2015 Nov 26}  & \multicolumn{3}{c}{\textit{Focal plane
    temperature drop}} \\[3pt]
\texttt{20160209} &   0730--5000 & 1\farcs25 & 1.07 \\[3pt]
\textit{2016 Feb 19}  & \multicolumn{3}{c}{\textit{Camera warmup \& corrector lens cleaning}} \\[3pt]
\texttt{20160223} &   1327--4845 & 1\farcs10 & 1.24 \\
\texttt{20160816} &   1900--5000 & 1\farcs08 & 1.06 \\
\enddata
\tablenotetext{a}{The local date at start of the night when the star flat
  exposures were taken or event occurred.}
\tablenotetext{b}{Median half-light diameter of the point spread
  function for the $i$-band exposures in the sequence.}
\tablenotetext{c}{$zY$ star flats were taken on the following night.}
\tablenotetext{d}{$zY$ star flats were taken on 10 Nov.}
\label{starflats}
\end{deluxetable}

\section{The DECam astrometric model}
\label{model}
Our goal is to produce an astrometric model that maps the $\vx^p$ of a
source to ICRS sky coordinates, such that any \emph{coherent} errors
are at $\lesssim1$~mas RMS.  Coherence applies here to both time and
space, meaning that the error should persist across more than one star
and more than one exposure.  Note that we are \emph{not} attempting to
model the following effects:
\begin{itemize}
\item Shifts in the centroids of individual detector pixels due to
  variation in lithography of the gate structures.  There is not
  enough on-sky stellar data to calibrate this for the 500 megapixels
  in DECam.  But astrometric errors
  due to pixel-to-pixel variations will behave as noise in individual
  stars' positions, and will not correlate between stars.  In
  well-designed use of DECam, a given star will be exposed on
  different parts of the array in each exposure, and hence this error
  will not correlate across time either.  In any case, the RMS
  variation in DECam pixel sizes is estimated (from flat-field
  behavior) to be at a few parts per thousand, or $<1$~mas.  Stellar
  position errors will be even lower since they average over a PSF
  containing $O(10)$ pixels.
\item Stochastic atmospheric distortions on $<1\arcdeg$ scale.  Such
  distortions are not coherent between exposures, but they are $\gg
  1$~mas and dominate the astrometric error budget for high-$S/N$
  detections.  The characteristics of these fluctuations are
  investigated in Section~\ref{atmosphere}.
\item Other sub-mas effects.
\end{itemize}

The DECam astrometric model was constructed through careful
examination of the residual astrometric errors in the star flat data.
The final choice of model is given in Table~\ref{mapelements}.  Here
we describe each element of the model in more detail, tracing
backwards from the collected charge in the pixel well back to the
top of the atmosphere.

\begin{deluxetable}{lccc}
\tablewidth{0pt}
\tablecolumns{8}
\tablecaption{Components of the DECam astrometric model}
\tablehead{
\colhead{Description} &
\colhead{Name} &
\colhead{Type} &
\colhead{Max.\ Size}
}
\startdata
Tree ring distortion &
$\langle\textit{band}\rangle\texttt{/}\langle\textit{device}\rangle\texttt{/rings}$
& \texttt{Template} (radial)  & $\approx0\farcs05$ \\
Serial edge distortion &
$\langle\textit{band}\rangle\texttt{/}\langle\textit{device}\rangle\texttt{/lowedge}$
& \texttt{Template} (X) & $0\farcs03$ \\
Serial edge distortion &
$\langle\textit{band}\rangle\texttt{/}\langle\textit{device}\rangle\texttt{/highedge}$
& \texttt{Template} (X) & $0\farcs03$ \\
Optics &
$\langle\textit{band}\rangle\texttt{/}\langle\textit{device}\rangle\texttt{/poly}$
& \texttt{Polynomial} (order$=4$) & $\gg1\arcsec$ \\
Lateral color\tablenotemark{a} &
$\langle\textit{band}\rangle\texttt{/}\langle\textit{device}\rangle\texttt{/color}$
& \texttt{Color}$\times$\texttt{Linear} & $\approx0\farcs04$ \\
CCD shift &
$\langle\textit{epoch}\rangle\texttt{/}\langle\textit{device}\rangle\texttt{/ccdshift}$
& \texttt{Linear} & $\approx 0\farcs1$ \\
Exposure & $\langle\textit{exposure}\rangle$ & \texttt{Linear} &
$\gg1\arcsec$ \\
Differential chromatic refraction & $\langle\textit{exposure}\rangle\texttt{/dcr}$ &
\texttt{Color}$\times$\texttt{Constant} & $\approx0\farcs05$ \\
\enddata
\tablenotetext{a}{The lateral color correction is set to \texttt{Identity}
  transformation for $izY$ bands.}
\label{mapelements}
\end{deluxetable}

\subsection{Tree rings}
\begin{figure}
\center
\includegraphics[width=0.9\columnwidth]{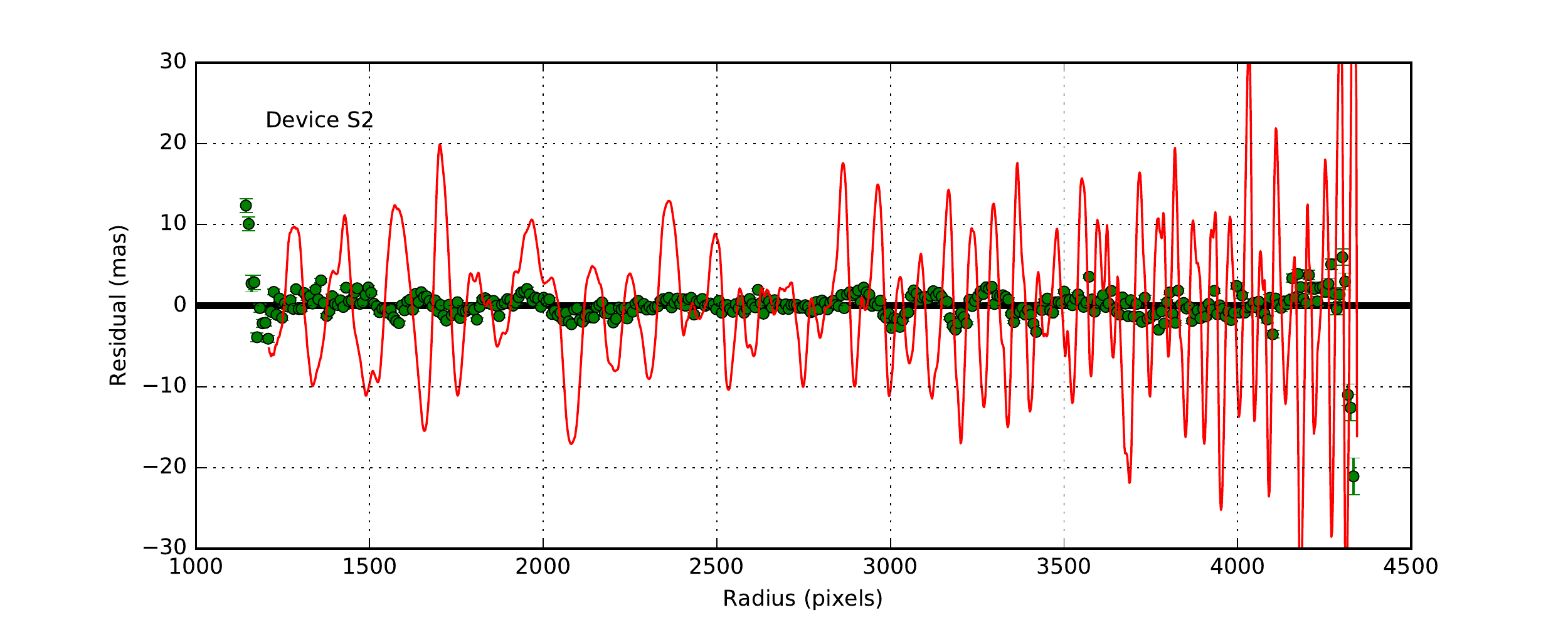}
\caption[]{\small The red curve is the tree-ring astrometric displacement
  template derived from the $r$-band flat-field image of CCD S2.  A
  spline-smoothed fit to the template is subtracted during its
  production to isolate the oscillatory portion that is due to the
  doping variations in the silicon boule.  The green points and error
  bars plot the binned astrometric residuals for star-flat detections
  in $gri$ bands on this detector.  The azimuthally-averaged
  astrometric residual has been reduced to $\approx 1$~mas~RMS.}
\label{treerings}
\end{figure}

 In $g,r,$ and $i$ bands, photons generate holes near the DECam CCD
surface and then have to drift the 250~$\mu$m thickness of the device
before being collected in the pixels. As described in \citet{plazas},
any electric field components transverse to the surface will cause the
charge carriers to drift sideways before collection and induce an apparent
astrometric shift.  The DECam CCDs are known to have two significant
sources of such stray fields.  The first are ``tree rings,'' which
arise from fluctuations in the impurity density of the silicon boules
from which the CCD wafers were cut.  The zone refining of the boules
results in approximate circular symmetry about the boule axis, and the
wafers are cut perpendicular to this axis, so the astrometric
distortions are realized as an irregularly oscillating pattern of
rings.  For some DECam devices, the ring centers are on the device,
for others the centers are off their edges.  Because the distortions
also produce oscillations in the solid angle of sky received by each
pixel, they are readily apparent in the flat-field images.  The
nearly-circularly-symmetric pattern in the flat fields implies that
the astrometric distortions share this symmetry and are directed
radially toward (or away from) the ring center.
As described in \citet{plazas}, we locate the ring center for each CCD by
visual inspection of the flat-field images, and then create templates of the expected
astrometric distortion about this center from a high-pass-filtered,
azimuthally averaged profile of the flat-field signal.
Figure~\ref{treerings} plots the template derived for a representative
device.

\begin{figure}[t]
\center
\includegraphics[width=0.9\columnwidth]{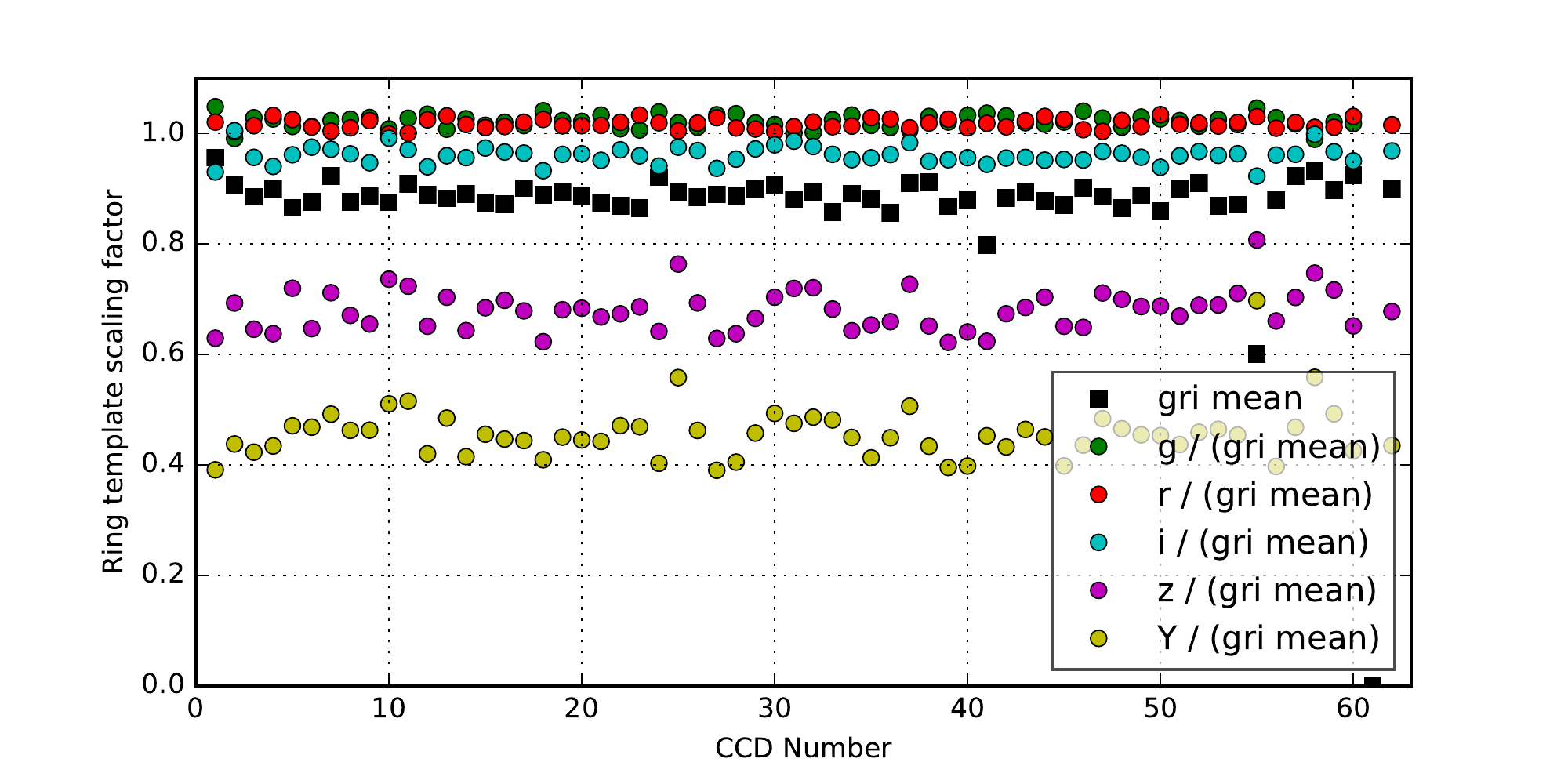}
\caption[]{\small The best-fit coefficients for the tree-ring
  distortion templates are plotted vs filter for all of the functional
  CCDs.  The coefficients are applied to the astrometric tree-ring pattern
  predicted from the $r$-band flat-field photometric rings.
  The black squares plot the mean, for each CCD, of the astrometric
  coefficients of the $g, r$, and
  $i$ bands.  The distortion is seen to
  be less than predicted by the flat-field templates, which is not understood.
  The other symbols show each filter's distortion amplitude relative to the
  $gri$ mean for that CCD.  These values decrease for $i,z$ and $Y$ bands, as
  expected due their longer absorption length in silicon. }
\label{ringfactors}
\end{figure}

In \wcsfit, the tree ring signal is realized as a \texttt{Template}
map, with both the variation and the displacement expected to be
purely radial to the rings.  We have a single free parameter for each
device/filter combination, which is a multiplicative scaling of the
distortion predicted by the template.  We do not allow for any time
variation of the tree ring 
signal, since the effect is literally built into the device.  We do
allow for a \edit{band} dependence, however, since photons in the $z$
and $Y$ bands penetrate well into the device and are therefore
expected to suffer less deflection before collection, on average.
Figure~\ref{ringfactors} plots the best-fit template coefficients for
all devices and filters.  We do not know,
why the tree ring distortions are seen to be only 80--90\%
of the values predicted from the $r$-band flat-field images.
But the scaling of these coefficients with
filter band hews closely to the values calculated from the
absorption-vs-wavelength characteristics of
silicon. Figure~\ref{treerings} plots the azimuthally averaged
residual position for all detections from the $gri$ exposures of a
representative device.  The RMS of this residual is at our goal level
of $\approx 1$~mas.

\subsection{Edges}
The electric field in the CCD also develops a substantial transverse
component near the device edges.  The subsequent astrometric
distortion and pixel-size variation is readily apparent in the
flat-field images as a ``glowing edge.''  It
is found that the flat-field (photometric) edge behavior is not a good predictor of
the astrometric distortions, so we derive a template for edge behavior
entirely from the stellar astrometry.  We assume throughout that the
edge distortion is directed in the $x$ direction (parallel to the serial
register on the short edge of the device) and is constant along $y$ at
each edge.  We first fit the star
flat data to a model with a \texttt{Piecewise} displacement term with
a free node position every 8 pixels within 180~pixels of each edge.

Note that the 25 (15) pixels of the device nearest to the long (short)
edges are completely masked from analysis because the distortion is
too large.  Thus we do not have useful stellar centroids closer than
$\approx30$ pixels to the $x=1,x=2048$ boundaries.  Any nodal values
in these regions are unconstrained and ignored.  There are also
unusable nodal values near the locations of any defective columns on a
device. 

\begin{figure}[t]
\center
\includegraphics[width=0.5\columnwidth]{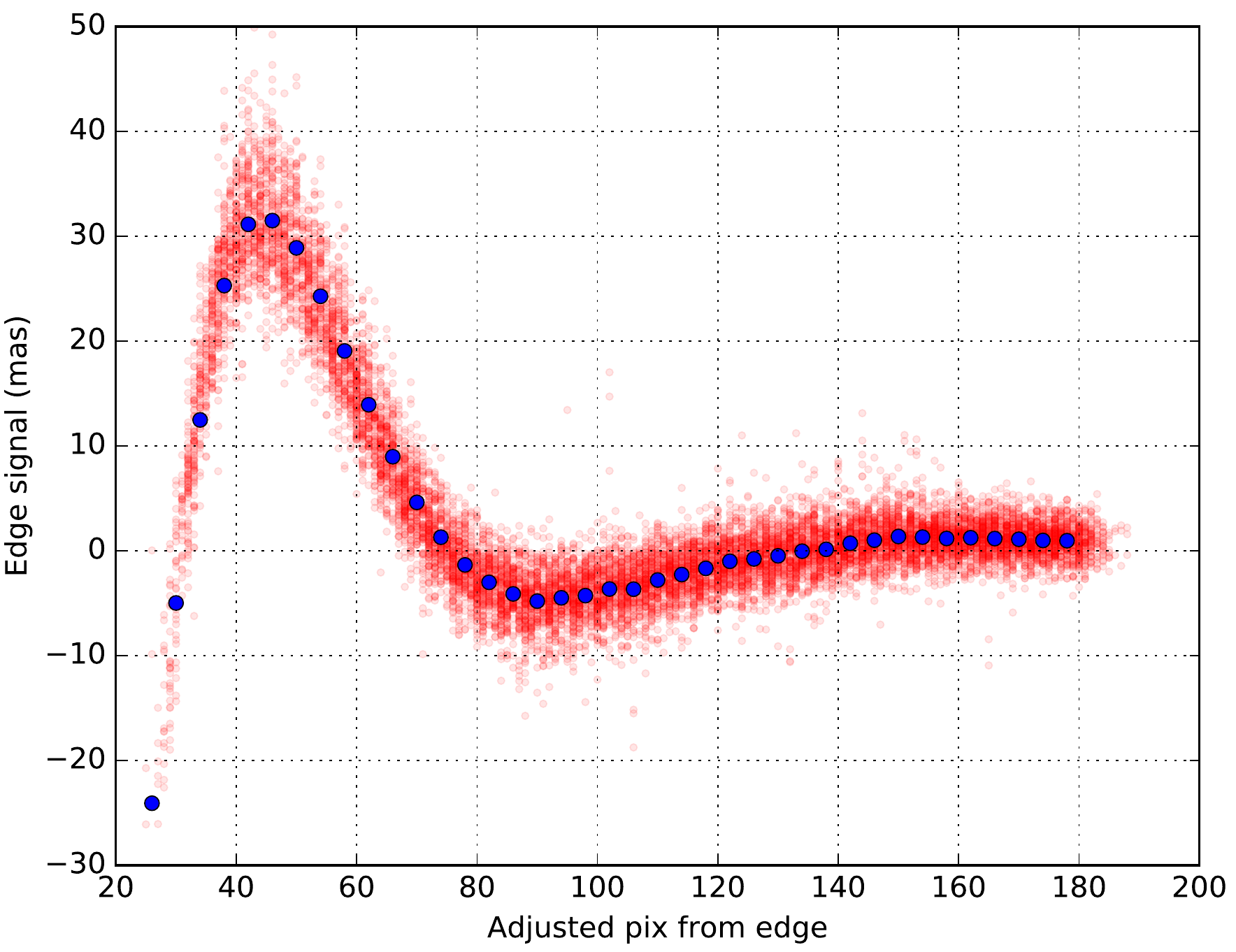}
\caption[]{\small The master template for the $x$ edge distortion is
  shown as the blue dots.  Each
  red dot is a binned astrometric residual for a single device/edge
  combination in a fit \emph{without} any modeling of the edge
  distortion, shifted horizontally by an amount judged to best align 
  with the master template.  The master template is derived from the
  median of all the red points, and is then used as a template for all
  $x$ edge distortions. The model interpolates linearly between the
  blue dots.}
\label{xedge}
\end{figure}

Upon examination of the best-fit piecewise solutions at the $x$ edges,
we find that all edges of all CCDs in all filters are consistent with a common
``master'' edge template, once we allow for a multiplicative scaling
and a shift as large as 12 pixels
(0.18~mm).   These shifts might from the finite precision of the
cutter tooling relative to the array during CCD dicing.  The master
edge template is shown in Figure~\ref{xedge}.  In
the final astrometric fit, we allow each device/filter combination to
have a \texttt{Template} pixel map at the high- and low-$x$ edges.
The templates are shifted versions of the master template, and the
scaling is left as a parameter for \wcsfit\ to optimize.

\begin{figure}
\center\includegraphics[width=0.6\columnwidth]{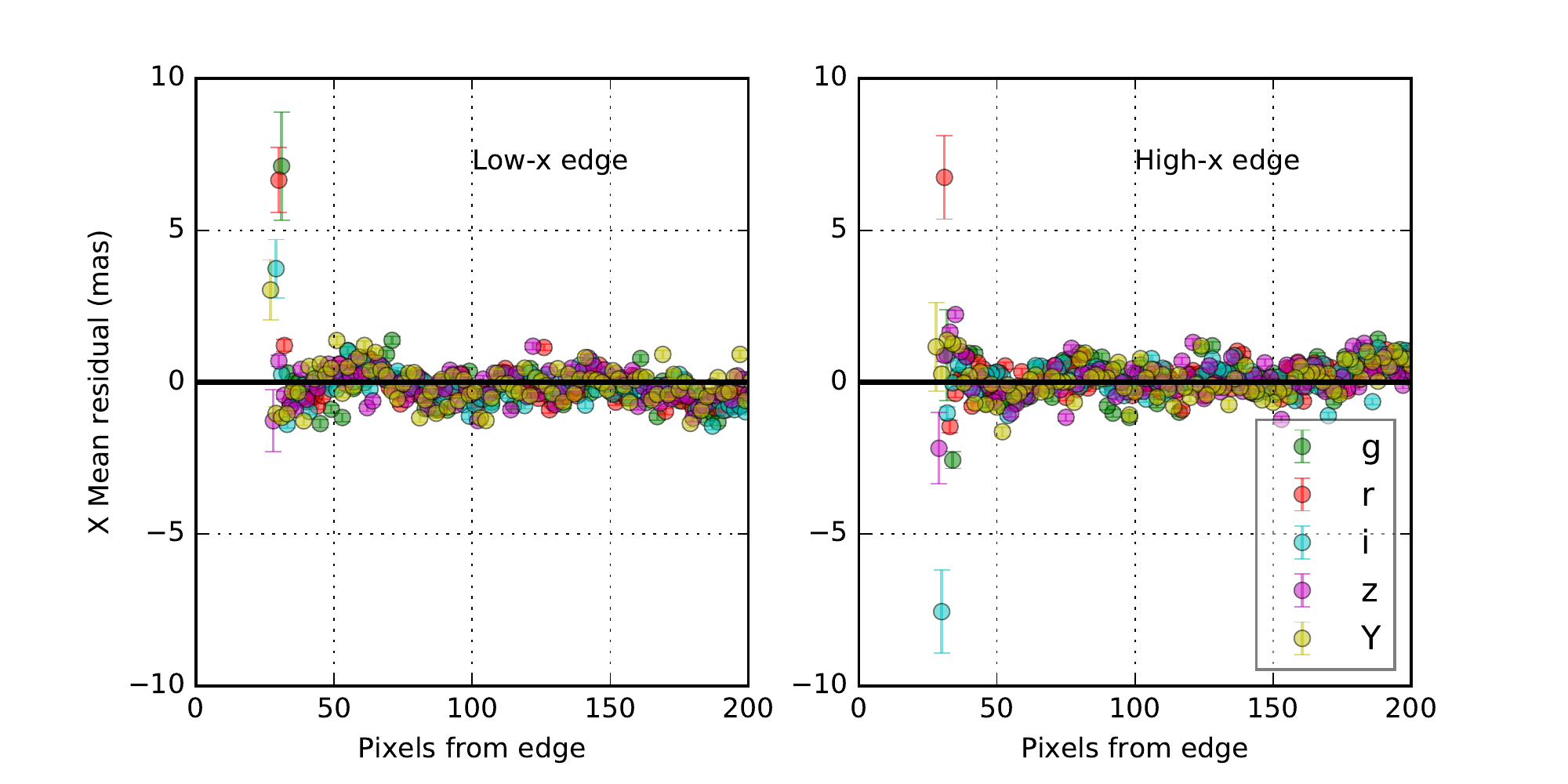}
\center\includegraphics[width=0.6\columnwidth]{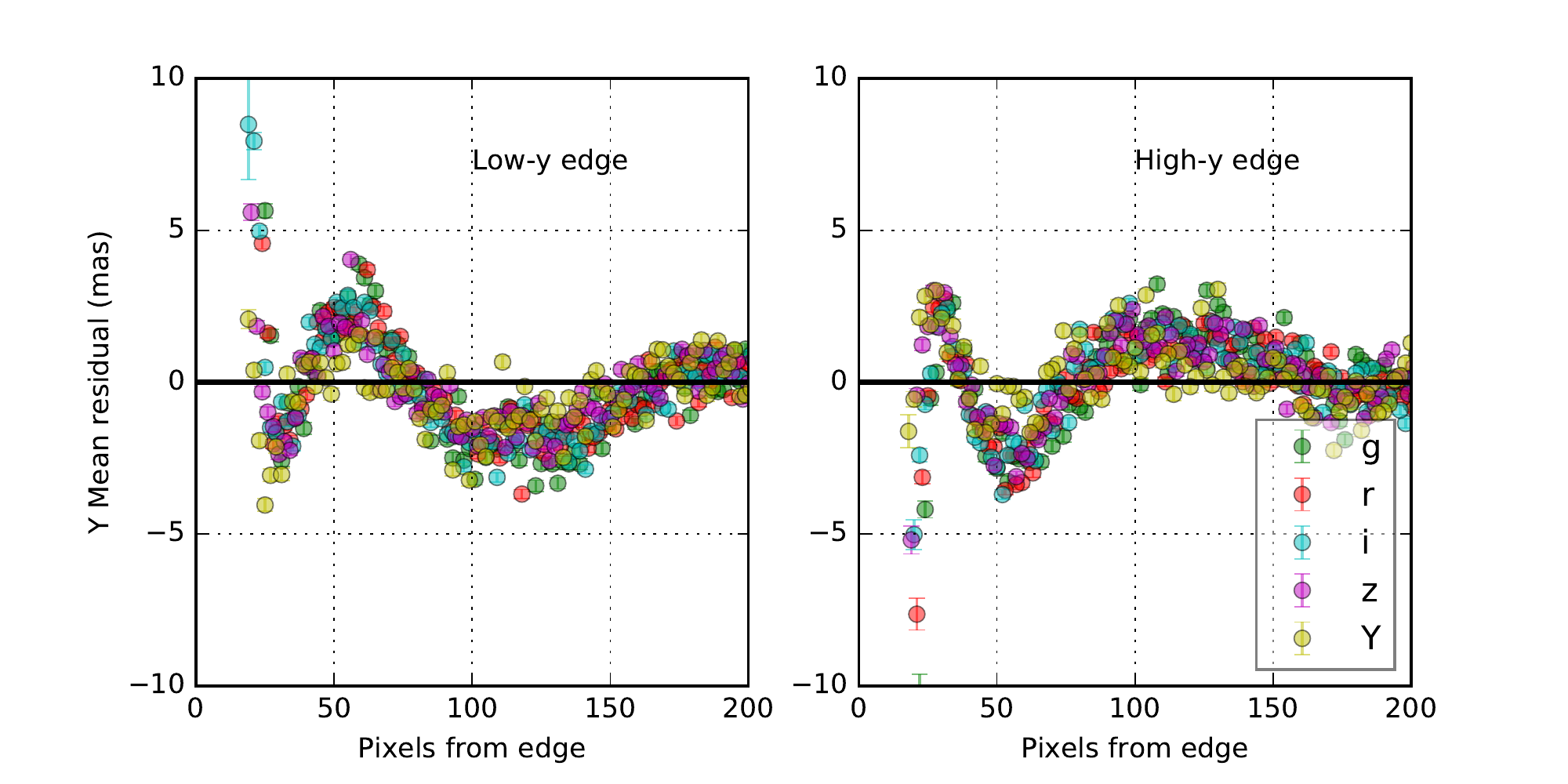}
\caption[]{\small Each panel shows the mean residual distortion vs distance
  from the CCD edge, after our final \wcsfit\ model which includes a
  multiple of the master template for the $x$ edges.  In each case we
  are averaging the displacement component perpendicular to the CCD
  edge.  The master template is seen to reduce RMS $x$ edge residuals
  to $\ll 1$~mas.  We have not implemented a correction for the $y$
  edges because the signal is $<3$~mas in all cases and affects only a
  few percent of the focal plane.  These plots average over all CCDs
  in the northern half of the array; the southern CCDs are installed
  with 180\arcdeg\ rotation and would swap the high and low sides.
  Plots for individual CCDs are consistent with noisier versions of
  the mean behavior.
}
\label{xyresids}
\end{figure}

Figures~\ref{xyresids} plot the binned displacement
residuals in all filters near all 4 edges of the CCD after the
$x$-edge template is included in the \wcsfit\ model.  The master
template reduces RMS $x$ residuals to well below 1~mas.  Note that we
have elected to make no correction at all for the glowing edge effect
on the short ($y$) edges, because the displacement is already $<3$~mas
before correction.  Since it affects only a small fraction of the
focal plane, the RMS error is $\ll1$~mas.

The best-fit coefficients to the master template are found to be
in the range 0.8--1.2 in the $gri$ bands, with lower values in $z$ and $Y$ as
expected again from the deeper photon conversion.  We take the edge
coefficients to be independent of time, as one would expect for such
detector-physics effects.

\subsection{Optics polynomials}
The vast majority of the nonlinearity in the $\vx^p\rightarrow\vx^w$
map is produced by the classical distortion of the optical system.  A
time-independent polynomial map for each filter/device combination,
with terms $x^m y^n$ up to order 
$m+n\le4,$ is used.   While the camera optics
have radial distortion at fifth (and higher) order, a
fourth-order-per-CCD solution
is found sufficient to capture the optical distortion, and the
placement of the CCDs in the focal plane. These polynomials carry 30
free parameters per CCD per filter, roughly 9000 for the whole array.
It is in this map that we change units from pixels (on the array) to
degrees (in the gnomonic projection about the telescope axis).

\subsection{CCD shifts}
The DECam astrometric map is observed to change over time by
$O(100)$~mas. We posit that these changes are dominated by small 
translations and rotations of the devices with respect to their
mounting plate, or other mechanical drifts.  We allow \wcsfit\ to
model this by adding a linear (affine) distortion, with 6 free
parameters, to each CCD, for every star flat epoch except the first
one.  The CCD shifts are taken to be identical in all filters.  The
results of these fits are examined in Section~\ref{stability}.

\subsection{Lateral color}
Any axisymmetric refractive optical system is expected to have color-dependent
radial distortion, leading to color terms described by odd-order
polynomials in radius.  We check this assumption by including in our
initial fits a more general color term:
a time-independent linear function of coordinates on each
CCD.  The displacement is assumed to be proportional to
\begin{equation}
c \equiv (g-i) - (g-i)_{\rm ref},
\end{equation}
where the reference color is chosen to be $0.44,$ the color in the natural DECam
$(g-i)$ system of the F8IV star C26202 from the Hubble Space Telescope
CalSpec
system.\footnote{http://www.stsci.edu/hst/observatory/crds/calspec.html}
We restrict the fit to stars with $-0.2\le (g-i)\le 1.8,$ to avoid M
stars for which the expected shifts may no longer be linear in $c$, and assume
that the color term is time-independent for a given filter.

\begin{figure}[ht]
\center
\includegraphics[width=0.55\columnwidth]{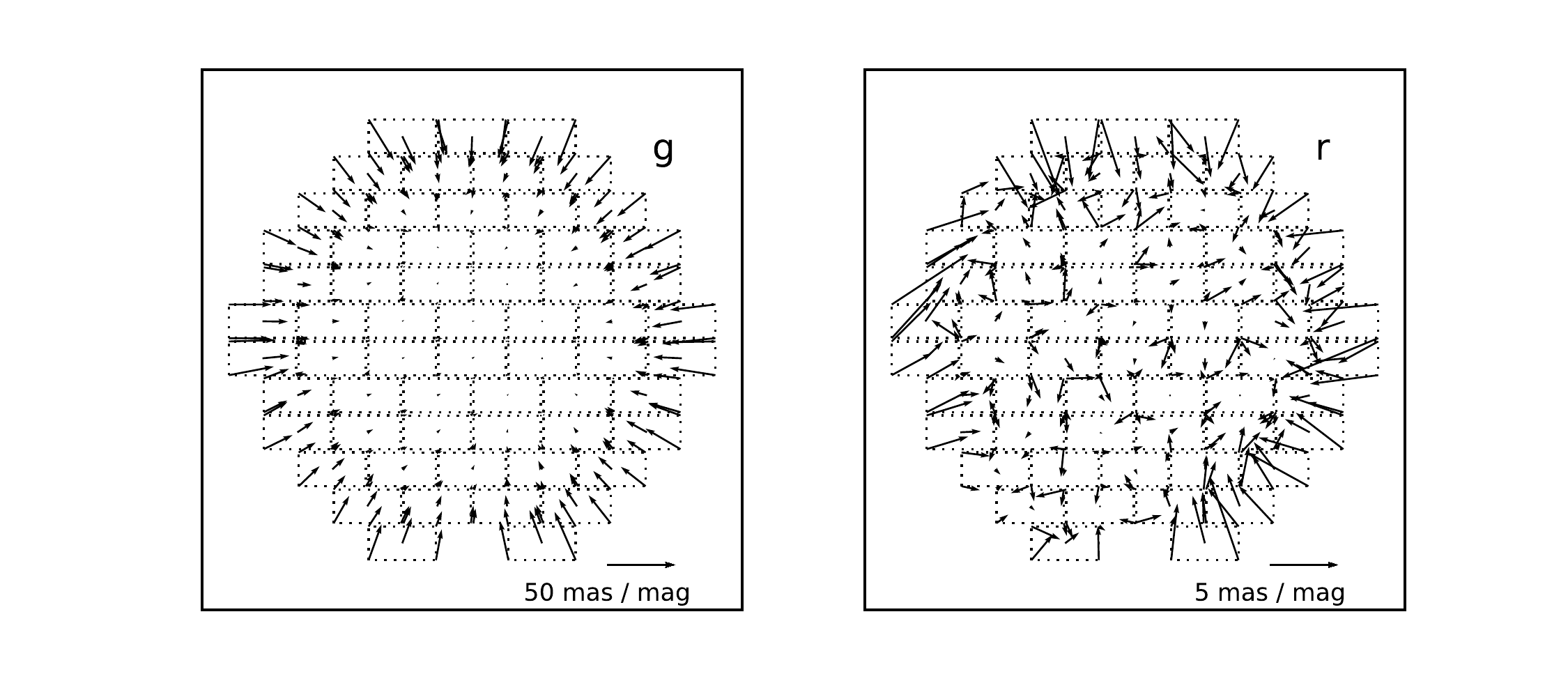}
\includegraphics[width=0.4\columnwidth]{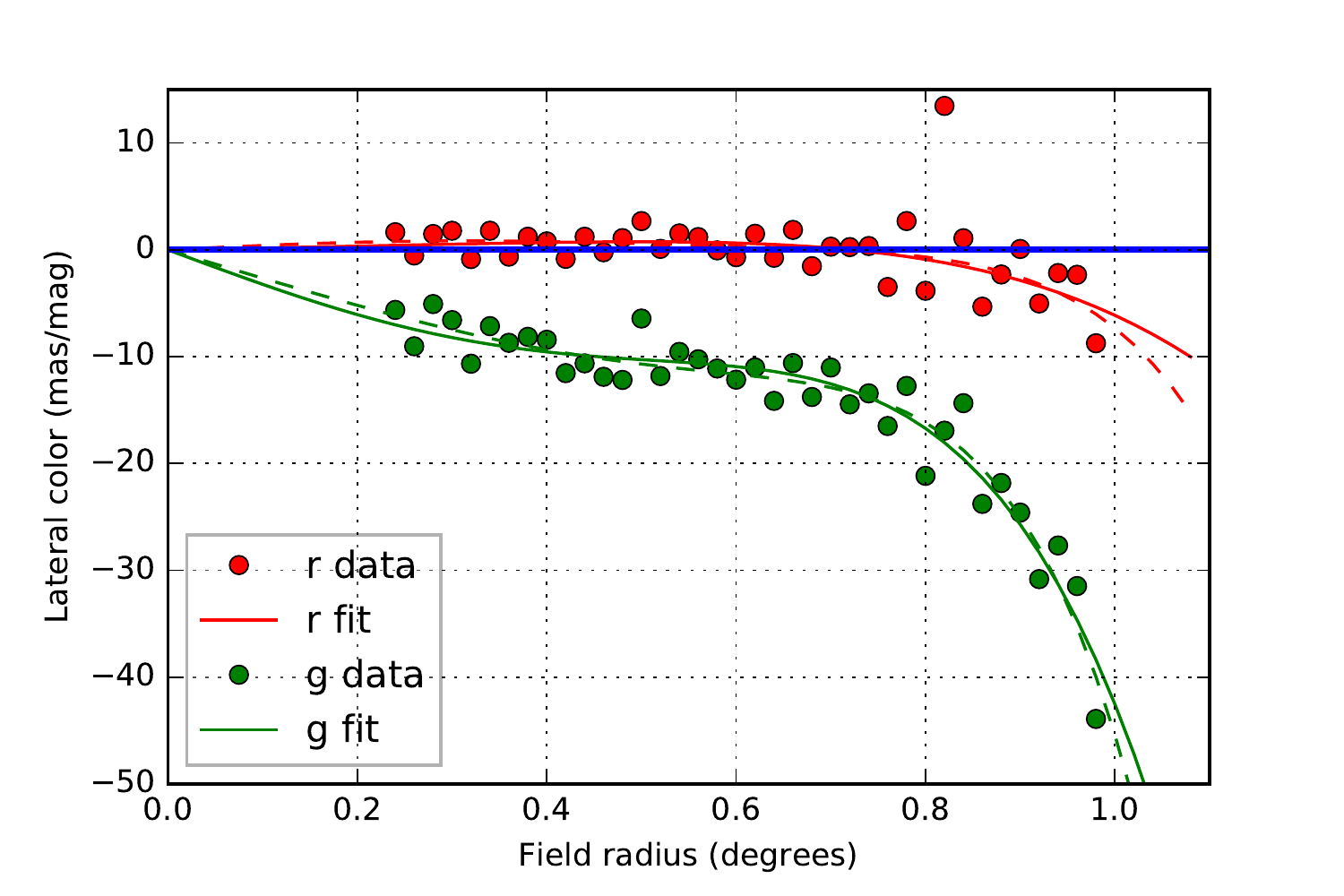}
\caption[]{\small The best-fit solutions for static color-dependent
  distortion are plotted for $g$ (left) and $r$ (center) bands.  Note
  the very different scales for each.  The dashed rectangles are the
  outlines of the 61 DECam CCDs that have been functional for at least
  part of this analysis.  The arrows plot the shift per magnitude
  of  $g-i$ color at the four corners of each CCD, using the best-fit
  model of linear dependence across each device.  As expected from the
  optical design, the pattern is radial, with barely detectable
  amplitude in $r$ band and no detectable $i, z$,or $Y$ band lateral
  color (not shown).  At right are the results of fitting the
  $g$ and $r$ data to a purely radial piecewise function of radius
  (circles).  We adopt as our final lateral color model the
  fifth-order polynomial fits to these models (solid lines), as a
  seventh-order fit (dashed lines) offers no significant improvement.
}
\label{lateral}
\end{figure}

Figure~\ref{lateral} plots the best-fit static color solution in the
$g$ and $r$ bands, which show the radial patterns (at mas accuracy)
and approximate amplitude
expected from the optical solution (S. Kent, private communication).
The solutions for $i, z,$ and $Y$ bands are, as expected, undetectably
small as the chromatic terms of the corrector lenses are weaker, 
and we disable their color terms for the final
\wcsfit\ run.  For the $g$ and $r$ bands, we continue by re-fitting to
a more restrictive function, namely a \texttt{Color} pixelmap wrapping
a radial \texttt{Piecewise} displacement.  
A fifth-order function of field radius is found to fit the resultant
piecewise function with 2~mas/mag RMS scatter (Figure~\ref{lateral}).
We used the fitted polynomials (implemented as  radial \texttt{Template} pixel
maps) for our final model, with no free parameters.

\subsection{Exposure solution}
Aside from the freedom to determine the pointing of the optic axis on
each exposure, we will clearly require the model to admit
exposure-to-exposure freedom to rotate, shear, and magnify the image
across the FOV, because these effects will be present (at many mas)
due to misalignment of the telescope equatorial mount,
temperature-induced variations in focal length, atmospheric
refraction, and stellar aberration from Earth's motion.  An active
optics system \citep{roodman} controls the position of the camera and
corrector with respect to the primary mirror; variations in this
position could also induce small time-dependent changes to the optical
distortion. We proceed with this linear freedom per exposure in our
analyses.  As noted earlier, atmospheric refraction should generate
quadratic terms at $O(10)$~mas; we will subsume these into our
investigation of stochastic atmospheric distortions in
Section~\ref{atmosphere}. 

\subsection{Differential chromatic refraction}
The color dependence of atmospheric refraction in the context of
wide-field cosmological surveys is studied by \citet{andresdcr} and
\citet{meyers}, who conclude that in some bands it will be present at
$O(10)$~mas and significant for weak gravitational lensing analyses.
The atmospheric refraction is very large ($1\arcmin\times\tan z$ at
zenith angle $z$) compared to our desired accuracy so the chromatic
effect is significant.  \edit{For a given site it is} expected to behave as
\begin{equation}
\label{dcreqn}
\Delta \vx^w = K_b  c \tan z \, {\bf \hat p}
\end{equation}
where $c$ is the object color (again $g-i$), 
${\bf \hat p}$ is the unit sky vector toward
the zenith (the parallactic angle), and $K_b$ is a constant derivable
for each band $b$
from the instrumental bandpass and atmospheric index of refraction.

\begin{figure}[t]
\center
\includegraphics[width=0.45\columnwidth]{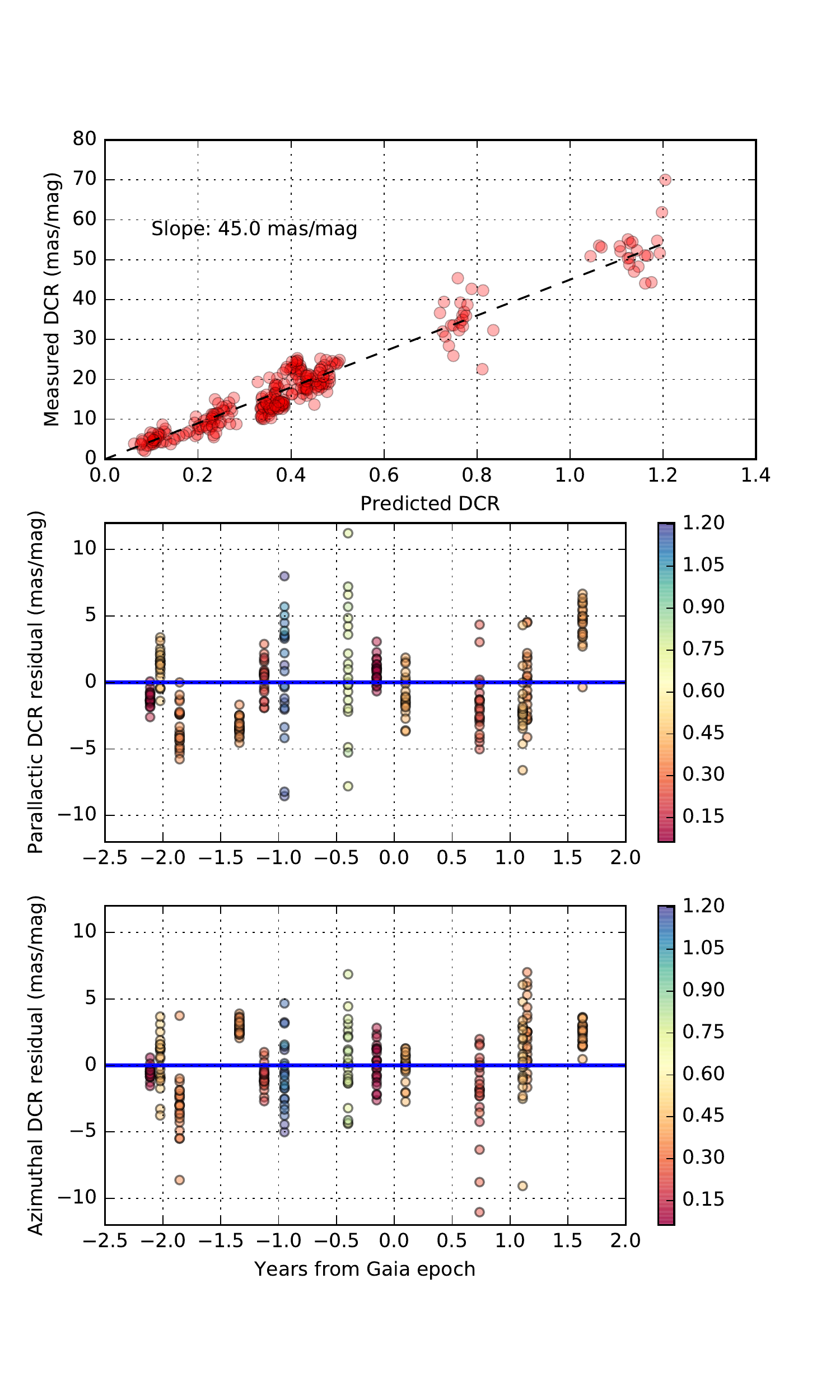}
\includegraphics[width=0.45\columnwidth]{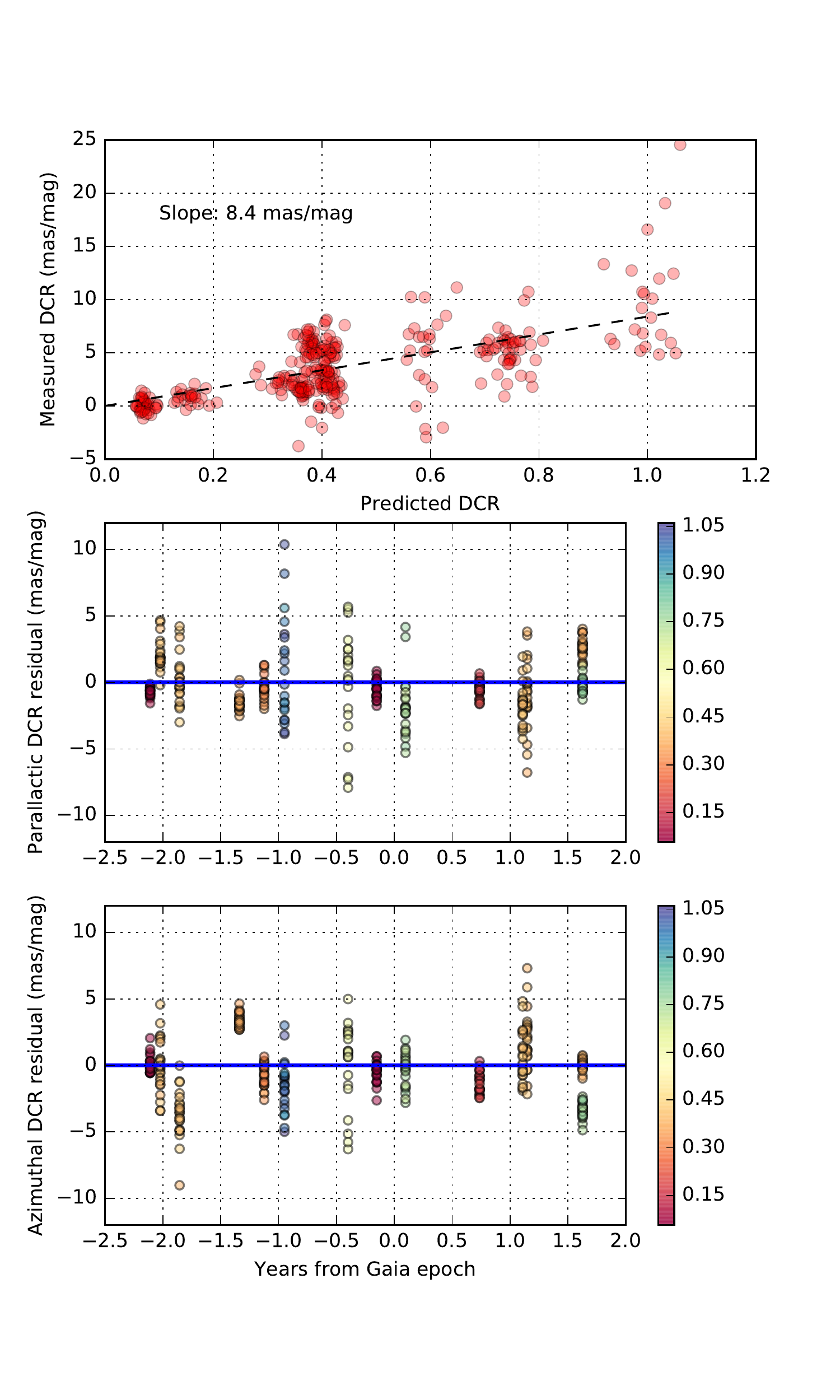}
\caption[]{\small The derived differential chromatic refraction (DCR) for the
  $g$ and $r$ star flat exposures (left and right columns,
  respectively) are compared in the top row to the predicted scaling
  with $\tan z$.  The top row plots the measured DCR component along
  the parallactic (zenithal) direction, with the dashed line showing
  the prediction \eqq{dcreqn} with the best-fit value of $K_b$ for
  each band $b$.  The middle row plots the residuals to \eqq{dcreqn}
  vs the date of the exposure, with the points color-coded according
  the value of $\tan z$.  The bottom row plots the azimuthal component
  of the measured DCR, which is expected to be zero.  The model works
  well, with 2--3~mas/mag RMS residuals, and no remnant trends with
  time or airmass.}
\label{dcrfig}
\end{figure}

We test this model by allowing each exposure to have
its own constant differential chromatic refraction (DCR) term $\Delta
\vx^w$ when fitting to the star flat detections.  Figure~\ref{dcrfig}
shows these results for $g$ and $r$ bands, along with the model
(\ref{dcreqn}) with the best-fit value of $K_b$.  The standard
atmospheric model is seen to describe the measured $c\,\Delta \vx^w$
well, with RMS residuals of 2--3~mas/mag.  Table~\ref{dcrtab} gives
the DCR amplitudes $K_b$ derived from the star flat data (supplemented
with the supernova field data described in Section~\ref{stability} for
the $i$ and $z$ bands).  The $K_b$ are in good agreement with the
predictions of \citet{andresdcr}, and for future use it should suffice
to simply fix the DCR term to \eqq{dcreqn} instead of allowing freedom
to each exposure.

\begin{deluxetable}{cc}
\tablewidth{0pt}
\tablecolumns{2}
\tablecaption{Differential Chromatic Refraction for DECam}
\tablehead{
\colhead{Band} &
\colhead{$K_b$ (mas/mag)}
}
\startdata
$g$ & 45.0 \\
$r$ & 8.4 \\
$i$ & 3.2 \\
$z$ & 1.4 \\
$Y$ & 1.1
\enddata
\label{dcrtab}
\end{deluxetable}

\section{\edit{Unmodelled} distortions}
\label{residsec}
\begin{figure}[ht]
\center
\includegraphics[width=0.22\columnwidth]{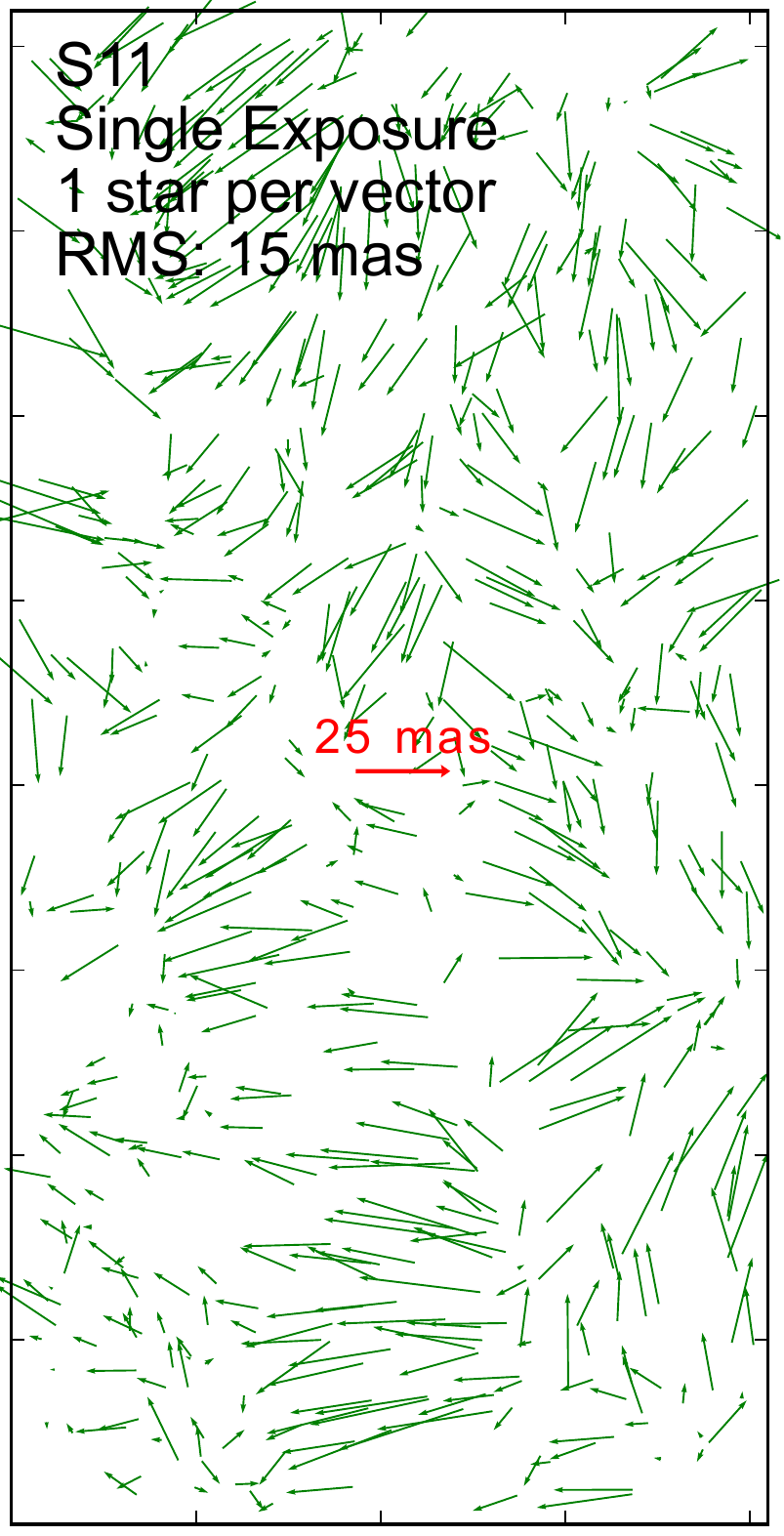}
\includegraphics[width=0.22\columnwidth]{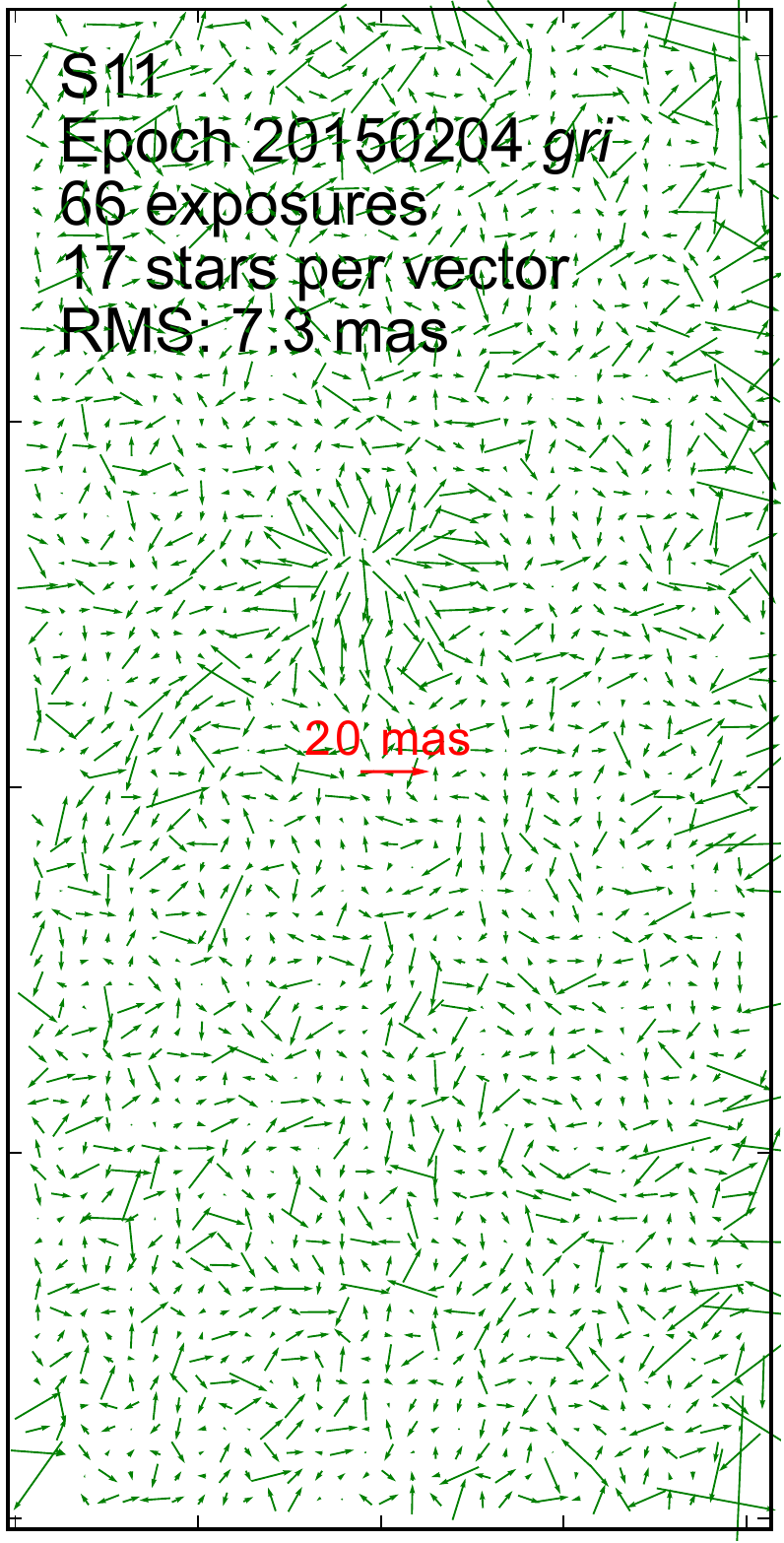}
\includegraphics[width=0.22\columnwidth]{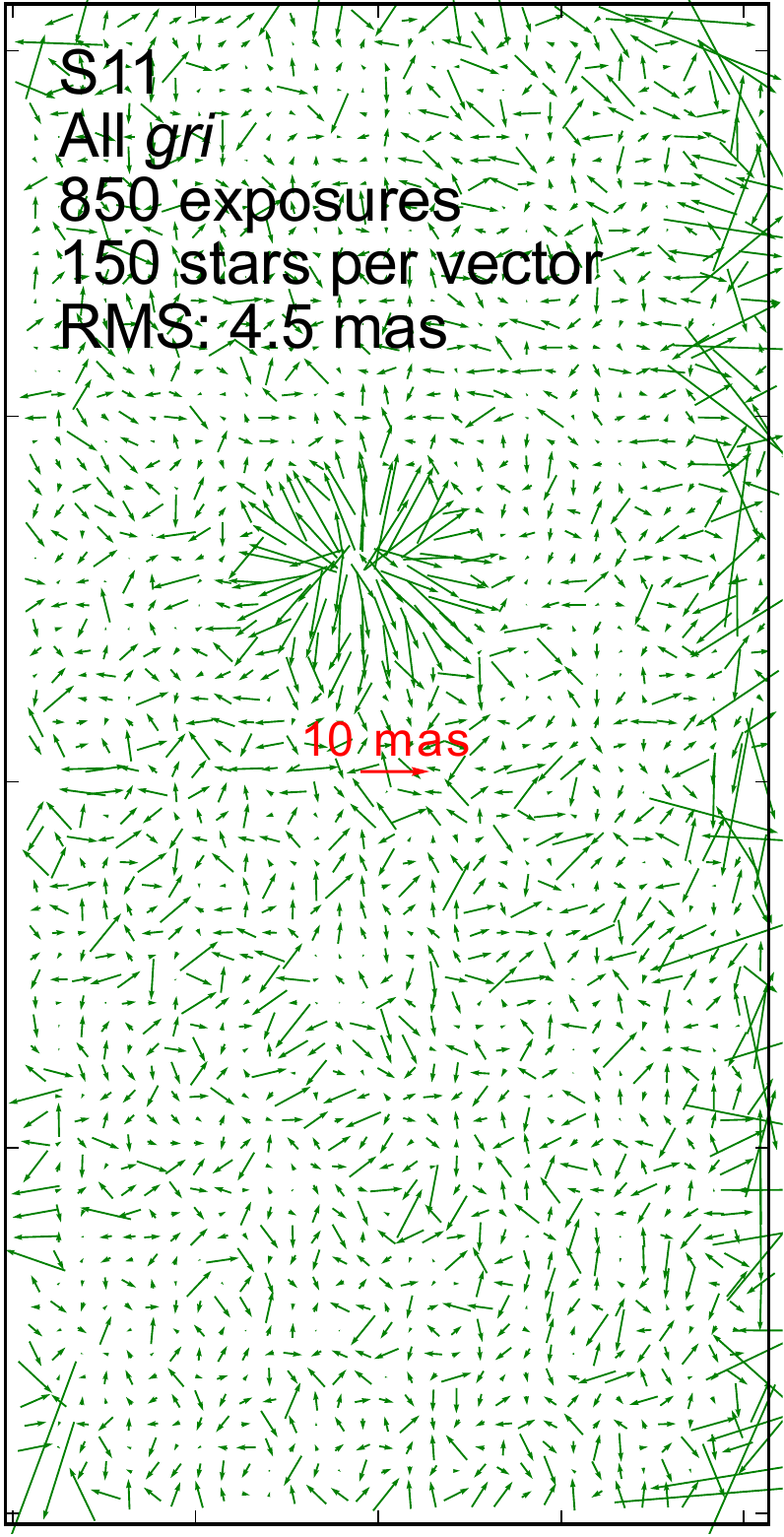}
\includegraphics[width=0.22\columnwidth]{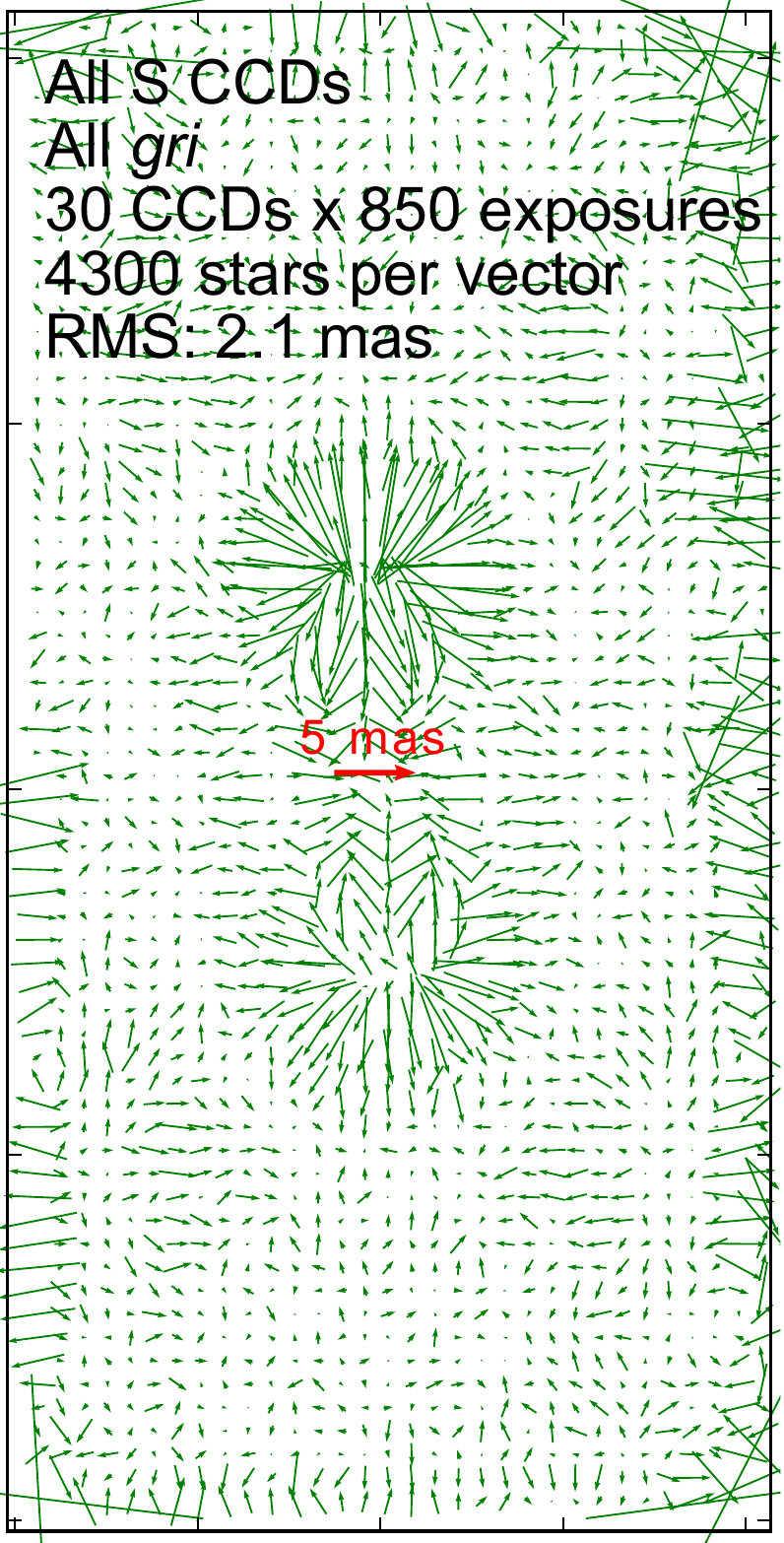}
\caption[]{\small Astrometric errors as a function of CCD position are
shown at left for all detections on a randomly selected detector (S11)
in a single exposure.  The pattern is dominated by atmospheric
turbulence.  Succeeding panels average larger sets of data in bins of
CCD position order to
reduce the atmospheric signal and reveal persistent errors in the
astrometric model.  Note the change of scale in each panel.}
\label{ccdresids}
\end{figure}

The first panel of Figure~\ref{ccdresids} plots the errors $\Delta\vx$
in the
stellar positions of a randomly selected single exposure on a randomly
selected CCD, relative to the mean positions
determined for the same stars from the entire stack of star-flat
exposures.  The residuals after application of our final astrometric
model are dominated by a coherent pattern with RMS
$|\Delta\vx|\approx15$~mas.  This pattern is found to differ from
exposure to exposure and is plausibly attributed to refraction by
atmospheric turbulence.  Before investigating these stochastic
distortions in Section~\ref{atmosphere}, we ask here whether there are
any distortion patterns that recur from exposure to exposure.  To find
them we will need to average down the stochastic atmospheric signal by
stacking and binning the residuals from many exposures.  

The middle panels of
Figure~\ref{ccdresids} show the result of averaging the residuals from
all of the $g,r,$
and $i$ band exposures from a single star flat epoch, and then from
all epochs.  The amplitude of the residual pattern drops steadily with
number of exposures included, although not as quickly as the square
root of the number of exposures, as would be expected if all remaining
errors were uncorrelated between exposures.  The all-$gri$ residuals
for this CCD exhibit clear coherence of several kinds:
\begin{itemize}
\item A radial pattern just above center contains distortions with
  amplitude of up to 30~mas.  This roughly coincides with the edge
  the electrical connector that is soldered and glued to the CCD, and
  protrudes through a hole in the CCD mounting board. The
  right-hand panel of Figure~\ref{ccdresids} averages over all 30
  functional DECam science CCDs mounted in the same orientation,
  showing that this pattern is recurrent and exists at both ends of
  the connector mount, and is likely a product of stresses induced in
  the CCD lattice the connector or the hole in the mounting board.
\item There are excess residuals along the long edges of the device,
  suggesting that the edge distortion is not uniform along the edge of
  the device.  The rightmost panel confirms that this is true in a
  systemic fashion for the devices.
\item The right-most panel shows the largest residuals in the corners
  of devices, plus two patches at the midpoints of the long edges.
  These six locations are known as ``tape bumps'' since they are
  underlain by thin spacers that define the thickness of the glue
  layer between the CCD
  and its carrier.  These regions, each 100--200 pixels on a side,
  exhibit structure in the flat fields that is indicative of stray
  transverse electric fields induced by lattice stresses.  There are
  clear astrometric disturbances associated with these fields as well.
  Because these are difficult to model and cover only a small fraction
  of the focal plan, we do not attempt to remove them: detections
  occurring on the tape bumps are flagged as having less reliable
  astrometry, and in fact have been omitted from the characterization
  and modeling performed in this paper.
\end{itemize} 

The DECam flat fields show evidence for modulations of the
pixel size with period $\Delta x = 27.33$~pixels, a behavior seen in
many CCDs due to the step-and-repeat accuracy of the mask generator
\citep{AK99}. 
A corresponding periodicity is detectable in the astrometric
deviations, but with peak-to-peak amplitude $<1$~mas, as is predicted
from the amplitude of the flat-field fluctuations.  We ignore this
effect for DECam.

\begin{figure}[t]
\center
\includegraphics[width=0.5\columnwidth]{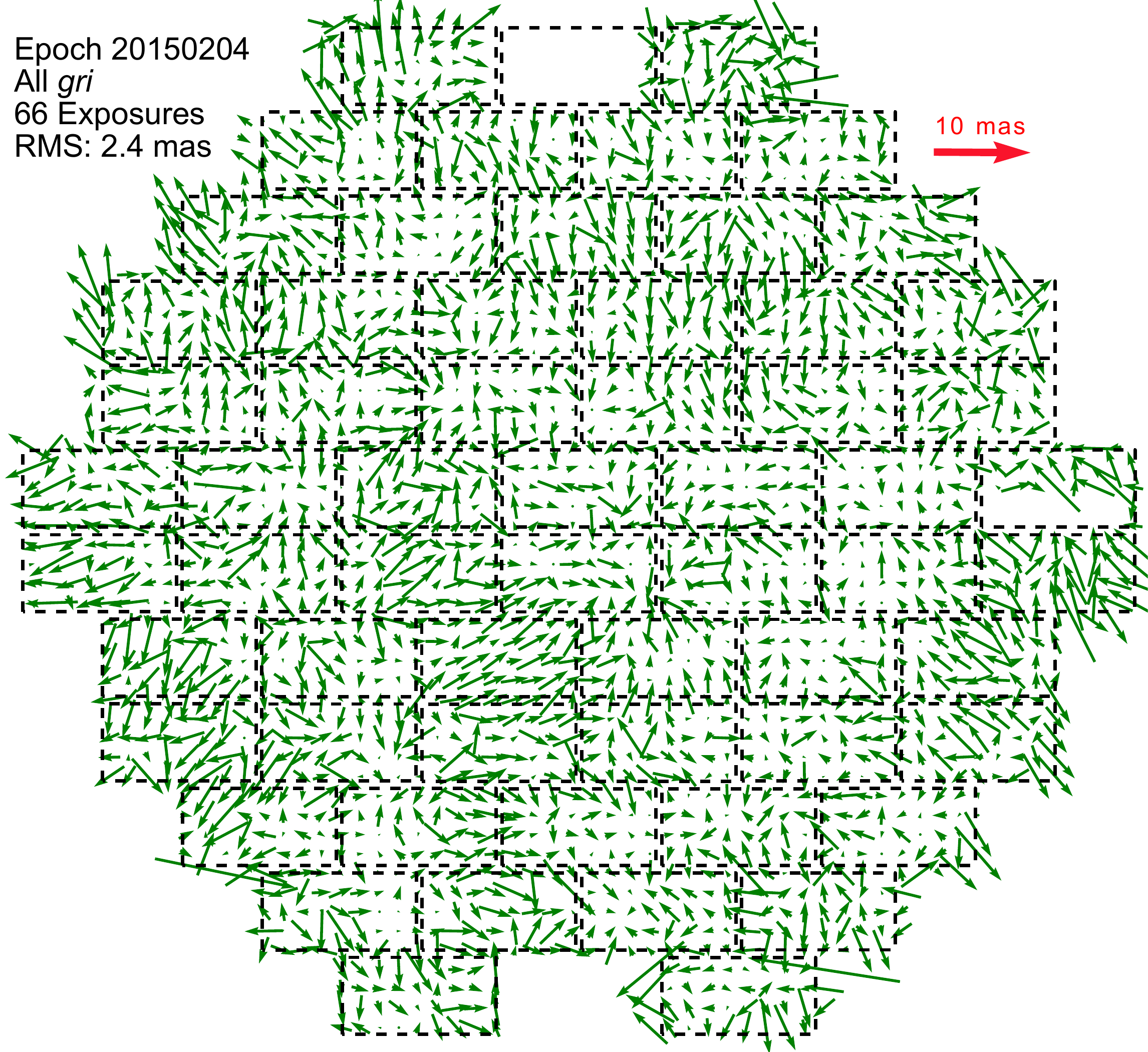}
\caption[]{\small The mean astrometric error of all detections in the
  $g,r,$ and $i$ bands during star flat epoch \texttt{20150204} is plotted vs
  array position.  No discontinuities at device boundaries are
  detectable (dashed boxes), and the signal is consistent with our model in which
  epoch-to-epoch changes are fully captured by linear adjustments to
  each CCD, plus the stochastic atmospheric signal. The DECam FOV is
  2\arcdeg\ in diameter, and the distortion field is magnified by
  $\approx50,000$ to conform to the red scale bar.}
\label{epochresids}
\end{figure}

A final question we address about the residuals to the model is
whether the linear-per-ccd \texttt{ccdshift} terms are sufficient to
describe the change of the astrometric solution between epochs.
Figure~\ref{epochresids} shows a test of this, whereby we plot the
mean residual astrometric errors for an entire epoch's $gri$
detections, averaged by position on the array.  The residuals are
consistent with the expectations of averaging 66 realizations of the
stochastic atmospheric pattern.  In particular there are no
statistically significant discontinuities across CCD boundaries.

We conclude that our astrometric model captures the recurrent
instrumental distortion pattern to an accuracy of 2--4~mas RMS.  The
only residual distortions that have detectable coherence between the
64-pixel (16\arcsec) bins of Figure~\ref{ccdresids} are those
associated with stresses from the CCD mounting, and with
inhomogeneities in the transverse electric fields generated at the CCD
edges. 

\section{Stochastic errors}
\label{atmosphere}
\begin{figure}[ht]
\center
\includegraphics[width=0.5\columnwidth]{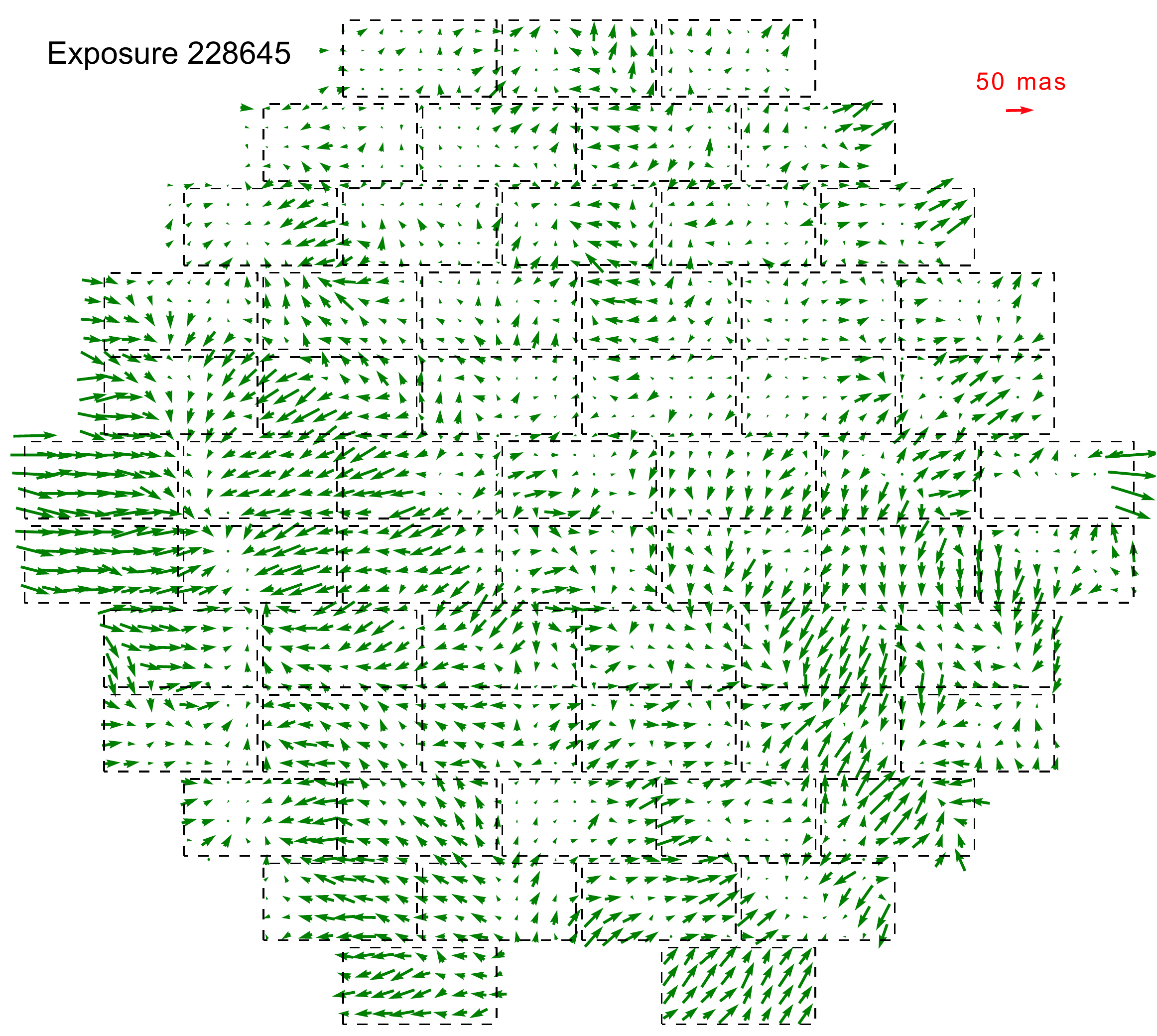}
\includegraphics[width=0.8\columnwidth]{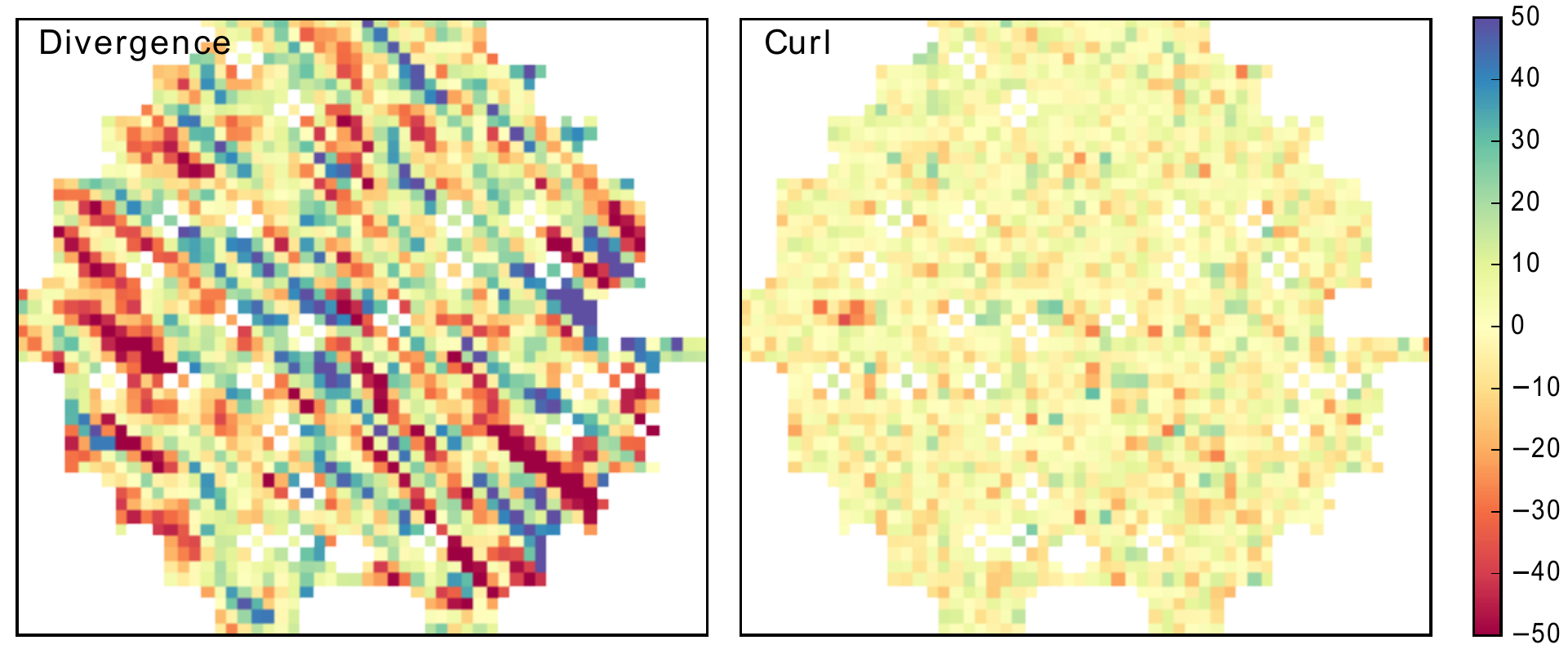}
\caption[]{\small At top are the astrometric residuals of detections
  in a representative exposure (228645, $z$ band), averaged in bins of
  focal-plane position.  Below are the divergence and curl of this
  vector field, plotted on a common scale.  The continuity of the
  vector field across chip boundaries, the curl-free nature of the
  field, and the streaky pattern of divergence strongly support the
  hypothesis that these distortions arise 
  from atmospheric turbulence.}
\label{atmos}
\end{figure}

After application of the astrometric model, the dominant form of
astrometric error is a field that varies from exposure to exposure and
has a coherence length of 5--10\arcmin.  Figure~\ref{atmos} plots the
residual vector field for a representative exposure.

\subsection{Atmospheric turbulence}
Multiple lines of evidence support the hypothesis that these
distortions arise from refraction by atmospheric turbulence:
\begin{itemize}
\item The patterns are uncorrelated between exposures, and thus change
  on time scales of 1 minute or less.  The only physical conditions
  that should change this quickly are the atmosphere and the settings
  of the hexapod that fixes the alignment of the camera to the primary.
\item The distortion pattern appears to be curl-free.  The lower panel
  of Figure~\ref{atmos} suggests that the curl arises from white
  noise, \ie\ errors in stellar positions due to shot noise.  This is
  shown more rigorously in Figure~\ref{ebcorr}, in which the
  2-point correlation function of astrometric errors is split into E- and
  B-mode components (curl- and divergence-free, respectively), as
  explained in Appendix~\ref{ebappendix}. The latter
  is seen to be consistent with zero.
  Curl-free distortion patterns are expected in the ray-optic limit,
  where the astrometric displacement of each photon is the gradient of
  the integral of the scalar index of refraction (time delay) along the line of
  sight to the star.
\item The distortion pattern is clearly anisotropic, with a long
  correlation length in one direction.  The preferred direction is
  roughly, but not exactly, consistent between exposures, as is
  expected from having the atmospheric turbulence pattern blowing
  across the field of view during the exposure. \edit{The streaky
    patterns are very similar to the maps of PSF ellipticity in
    short exposures with the CFHT Megacam presented by
    \citet{heymanspsf}, which they also attribute to wind.}
\item The power spectrum of the distortion in the cross-wind direction
  is roughly consistent with that expected of Kolmogorov turbulence.
\item The amplitude and correlation length of the distortion are
  roughly consistent with 
  numerical simulations of Kolmogorov turbulence (J. Peterson, private
  communication).  The simulations suggest that the astrometric
  perturbations are strongly dependent on the outer scale of the
  turbulence.
\end{itemize}
If the stochastic distortions do indeed arise from atmospheric
turbulence, we expect their amplitude to decrease with the square root
of exposure time as we average over phase screens.  We cannot verify
this with our data since nearly all star flat exposures were taken
with 30~s exposures, save the first three epochs which used 50~s.
While these early epochs do show the lowest stochastic distortion
(see Figure~\ref{astrohist}), there is substantial variation from
epoch to epoch so we cannot draw any quantitative conclusions.

\subsection{Behavior of the stochastic component}
\begin{figure}[t]
\center
\includegraphics[width=0.6\columnwidth]{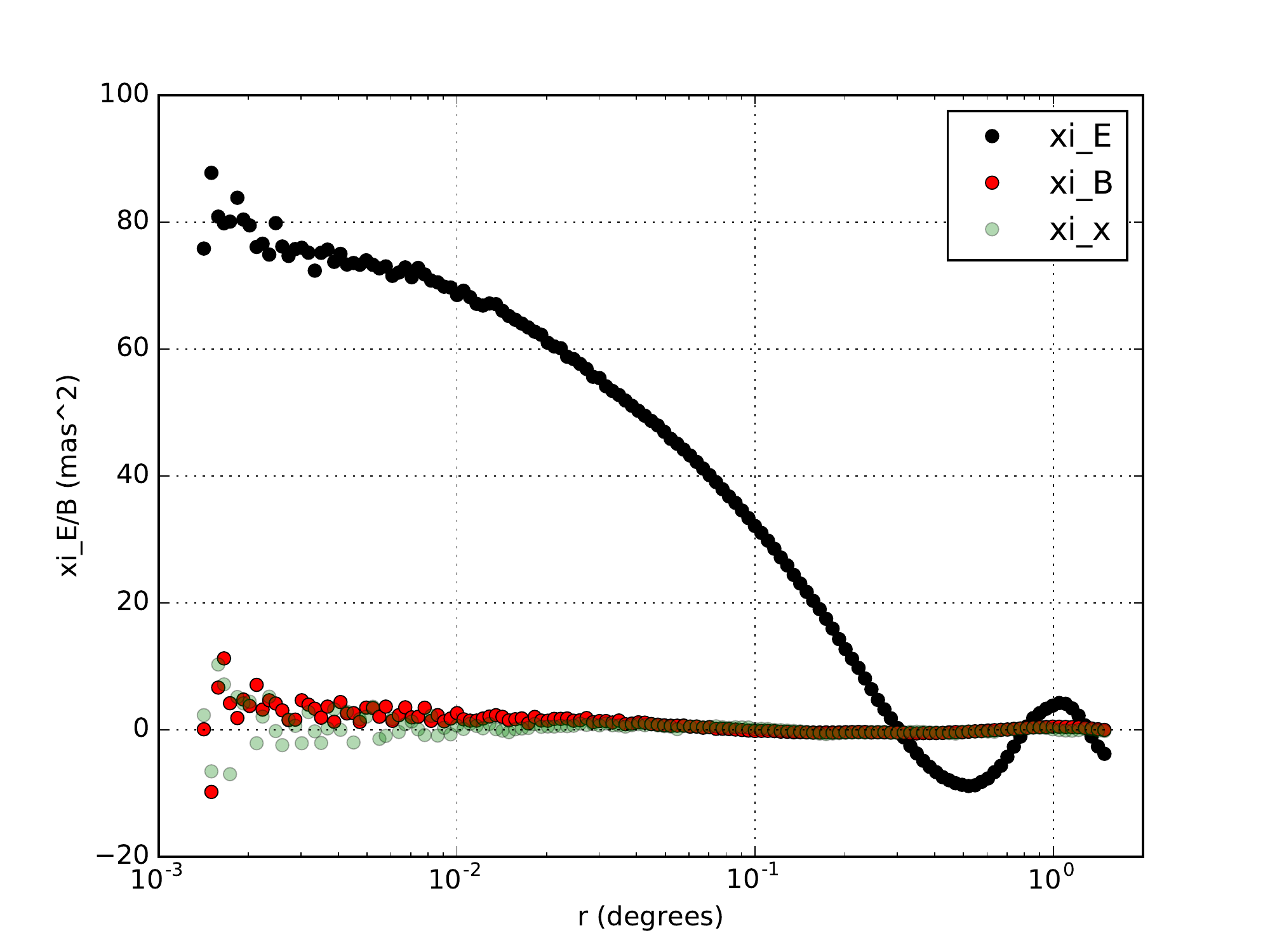}
\caption[]{\small Two-point correlation function $\xi(r)$ of the
  astrometric errors, averaged over 20 $z$-band exposures in the
  20121120 epoch, which exhibits the weakest stochastic distortion
  signal.  The signal is split into $\xi_E$ (the $\xi$ of the curl-free
  portion of the vector field), $\xi_B$ (divergence-free), and the
  cross-correlation $\xi_\times$ between these two.  As expected for
  any parity-invariant process, $\xi_\times$ is consistent with zero.
  Atmospheric refraction should have $\xi_B=0$, consistent with the
  observations.  The oscillations in $\xi_E$ at $r$ above
  $\frac{1}{4}$ of the field diameter are to be
  expected, since we have subtracted the best-fit cubic polynomial
  from the residual pattern.}
\label{ebcorr}
\end{figure}

For a more quantitative picture of the stochastic/atmospheric
distortion field, we produce its 2-point correlation function
\begin{equation}
\xi_+(r) \equiv \left\langle \Delta\vx_i \cdot \Delta\vx_j
\right\rangle_{|\vx_i-\vx_j|=r},
\end{equation}
where $\vx_i$ is the sky position of detection $i$, $\Delta\vx_i$ is
the measurement error on this position, and the average is taken over
all pairs of detections $i\ne j$ in the \emph{same exposure} separated
by distance $r$ (in practice the ``true'' position is taken as the average
of our many measured positions).
Appendix~\ref{ebappendix} explains how $\xi$ can be split into two
components $\xi_E$ and $\xi_B$ which arise from the curl- and
divergence-free parts of the vector field, respectively.  These are
plotted in Figure~\ref{ebcorr}, where it is clear that $\xi_B$ is
insignificant in comparison to $\xi_E,$ as expected for atmospheric
refraction.  In this case, and if we consider the turbulence to be a
Gaussian random field, then $\xi_E=\xi_+$ fully characterizes the
field.\footnote{A complete description would require inclusion of
  directional dependence of $\xi$ since the field is anisotropic.}

Before calculating $\xi$, we subtract from the $\Delta \vx_i$ the
best-fit cubic polynomial function of field coordinates.  As discussed
earlier, we expect significant linear and quadratic-dependence
distortions from normal (homogeneous) atmospheric refraction;
turbulent refraction should also have a substantial large-scale
component, and indeed we observe $\approx25\%$ of the distortion
variance to come from this polynomial signal.  Since the low-order
component is easily determined in practice by fitting to the Gaia
catalog, we remove it from our analysis, leaving small-scale
distortions.  Note that the virtue of using $\xi(r)$ is that it is
unaffected by the shot-noise measurement errors of the stellar
positions (for $r>0$) and therefore is a pure measure of the
astrometric map.

We characterize the astrometric correlation function by two
quantities, $\xi_0$ and $R_{\rm corr}.$  The former is essentially the
small-scale (largest) value of $\xi_+,$ defined as
\begin{align}
\xi_0 & \equiv  \frac{\int_0^\infty r\,dr\, W(r)
        \xi_+(r)}{\int_0^\infty r\,dr\, W(r)} \\
W(r) & \equiv e^{-r^2/2\sigma^2}, \qquad \sigma=1\arcmin.
\end{align}
The second quantity $R_{\rm corr}$ is defined as the radius at which
$\xi_+(R_{\rm corr}) = 0.5 \xi_0,$ \ie\ the smoothing scale that would
cut the astrometric variance in half.

\begin{figure}[t]
\center
\includegraphics[width=\columnwidth]{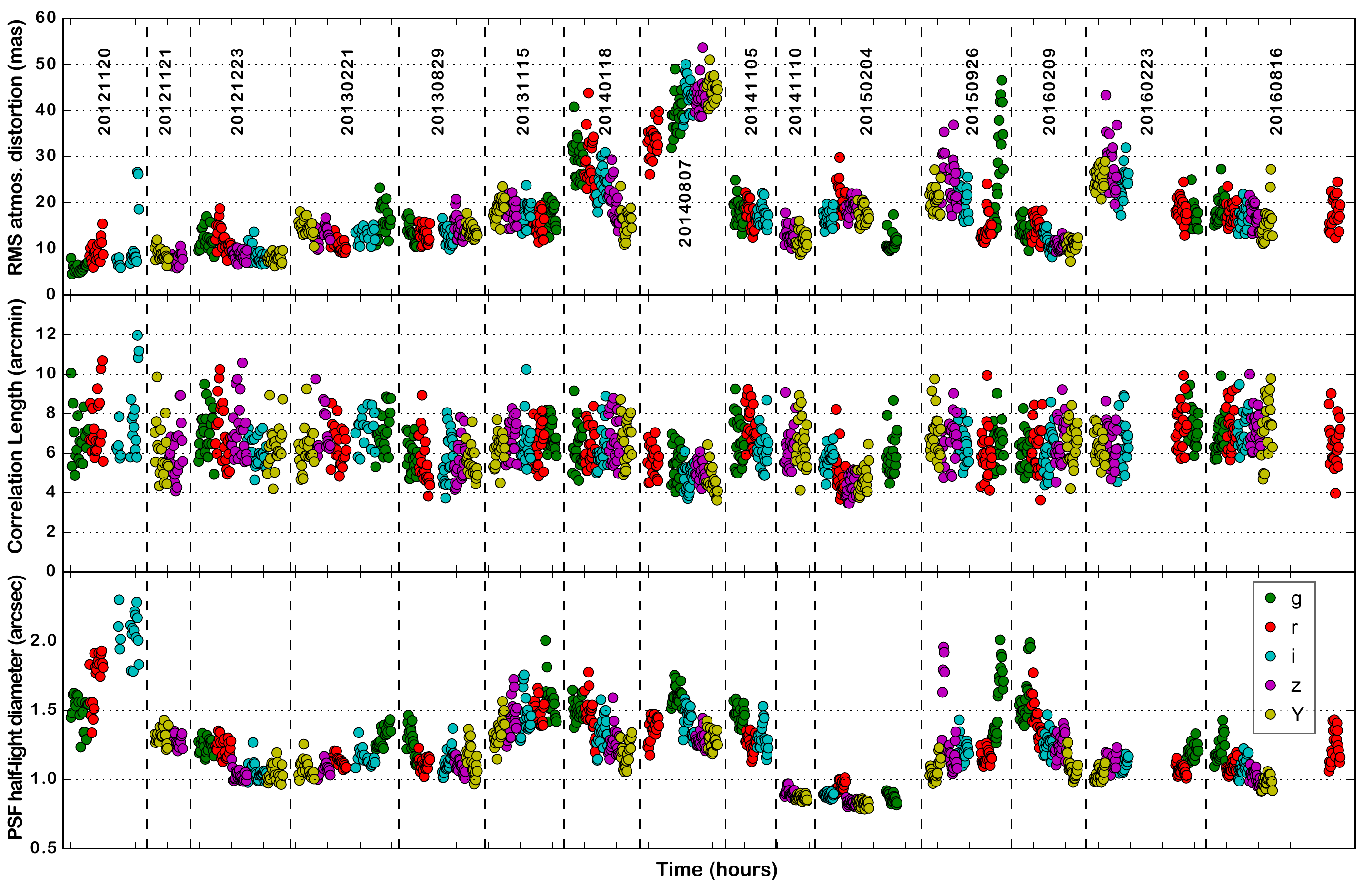}
\caption[]{\small The RMS stochastic astrometric distortion
  $\sqrt{\xi_0}$ (top), the correlation length $R_{\rm corr}$ of the
  distortion, and the half-light diameter $D_{50}$ of the PSF are
  plotted vs time for all star-flat exposures.  Each horizontal tick
  marks one hour, and the vertical lines represent the days to months
  between sets of star flat observations.  Epochs are labelled across
  the top.  The amplitude of astrometric distortion is only partially
  correlated with the seeing.}
\label{astrohist}
\end{figure}

Figure~\ref{astrohist} plots the values of $\xi_0$ and $R_{\rm corr}$
for all the star flat exposures under analysis, along with the
half-light diameter $D_{50}$ of the PSF in each exposure.
It is clear that there are nights when a degradation of seeing is
accompanied by an increase in astrometric distortion (\eg\ 20140118,
20150926), as one might expect if both are proportional to
the amplitude of a strictly Kolmogorov turbulence spectrum.  However
there are also cases of anti-correlation, and the mean seeing of a
night is a very weak predictor of astrometric accuracy.  In
particular, the epoch 20140807 is astrometrically awful, exhibiting
30--50~mas RMS atmospheric contribution whereas most other epochs are
10--20~mas RMS.  Yet the seeing on that night was not as poor.
Clearly there are other variables besides Kolmogorov amplitude,  such
as wind speed or outer scale, that determine the astrometric quality
of the night.

The correlation angle is in the range $4\arcmin< R_{\rm corr} <
10\arcmin$ at all epochs, with no apparent relation to the seeing.
This suggests that interpolation between stars in Gaia catalog, with
source density of $\approx 1$ star per arcmin$^2$ at high latitude,
could significantly reduce the stochastic atmospheric errors.  We
investigate this in Section~\ref{interpolation}.

The DECam measures of stochastic atmospheric astrometric fluctuations are
in rough agreement with previous characterizations by
\citet{hangatewood}, \citet{zacharias}, and \citet{bouy} (and
references therein), in terms of typical amplitude at good sites, and
substantial night-to-night variation that is at best tenuously tied to
the seeing FWHM.

\section{Solution stability}
\label{stability}
Is the DECam astrometric solution stable over weeks, days, or years?
We already know that there is short-term (seconds) variability due to
atmospheric turbulence, at a typical level of 10--20~mas RMS in a
30 s exposure and 5--10\arcmin\ correlation length.  But this
should average to zero with longer exposures.
We have verified in Section~\ref{residsec} that the astrometric
errors within a given star flat epoch (\ie\ a few hours' clock time)
are consistent with a single solution, up to the accuracy allowed by
the stochastic atmospheric distortions, as long as we allow for
expected exposure-to-exposure variations at low order across the focal plane.
We are
interested in the duration over which a single astrometric solution
can otherwise be considered to maintain few-mas accuracy.

\begin{figure}[tb]
\center
\includegraphics[width=\columnwidth]{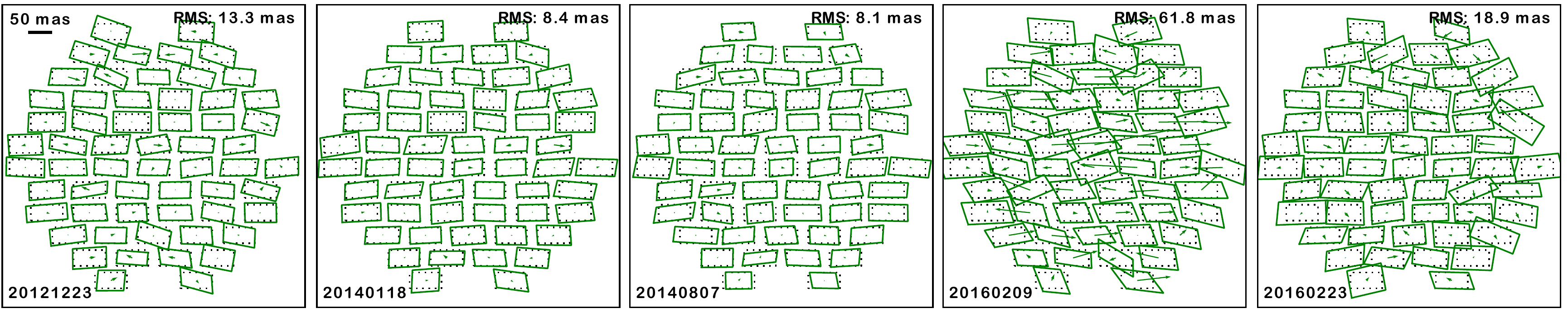}
\caption[]{\small The CCD shifts derived for a subset of the star flat
  epochs are shown, after removal of any exposure-wide cubic
  polynomial distortions.  In this and subsequent figures, the motion
  of the center of each CCD is indicated by the arrow, with scale
  shown in the upper left.  The green rhombi show the distortion of
  each CCDs shape, exaggerated such that the shift with respect
  to the (undistorted) black outlines depicts the distortion of the
  device at a scale corresponding to the bar. The epoch-to-epoch
  shifts can be large relative to the typical stochastic atmospheric
  distortions. Note that the last two epochs plotted are only 14 days
  apart but differ by $>100$~mas in places.}
\label{sfdrift}
\end{figure}
 
Our fit to all of the star flat data allowed for variations between
star flat epochs in the form of a free linear transformation for each
device.  Figure~\ref{sfdrift} depicts the ``CCD shift'' patterns found
to best fit 5 of the epochs.  In these plots, and in all analysis, we
removed from the CCD shift coefficients any components consistent with
an overall cubic polynomial distortion of the focal plane, since we
know that the solution will have time variability of this nature which
must be resolved on an exposure-by-exposure basis, \eg\ by using Gaia
reference stars.  The Figure makes it clear that there are
epoch-to-epoch changes that would dominate the stochastic errors even
in a single 30 s exposure.  The relative motions of CCDs can be
surprisingly large, \eg\ over 100~mas, or $>6$~$\mu$m in the focal plane,
over a time period of just 14 days, in the case of the last two epochs plotted.

The star flat sequences were taken too infrequently to resolve the
temporal behavior of the CCD shifts.  Fortunately the \textit{Dark
  Energy Survey} observing program includes repeated visits
to 10 fields in a search for high-redshift supernovae.  Each field is
imaged roughly once per week during the 6-month DES observing season.
We examine here the stellar detections in $\approx1500$ exposures
taken of the SNC3 field in the 4 years following camera
commissioning.  Visits to this field usually comprise $3\times200$~s
exposures in $g$ band, $3\times400$~s in $r$, $5\times360$~s in $i$,
and $11\times330$~s in $z.$  After matching all of the stellar
detections in these images, we run \wcsfit\ on the 1123 $i$ and $z$
band exposures from 112 distinct nights during which the photometric
solutions indicate absence of clouds and the seeing $D_{50}$ is predominantly
$<1\farcs6.$  This \wcsfit\ adopts the astrometric solution
derived from the star flats, holding all parameters fixed except:
\begin{itemize}
\item the linear solution for each exposure,
\item the differential chromatic refraction constant for each
  exposure, and
\item the linear CCD shifts, one per CCD per epoch (night) of
  observations.
\end{itemize}
From this solution we extract the CCD shifts for each SNC3 epoch,
combining them with the CCD shifts for the star flat data, and
projecting out a FOV-wide cubic polynomial fit to each epoch.  We
analyze only the 59 CCDs that are fully functional over the 4 years.

\begin{figure}[ht!]
\center
\includegraphics[width=0.8\columnwidth]{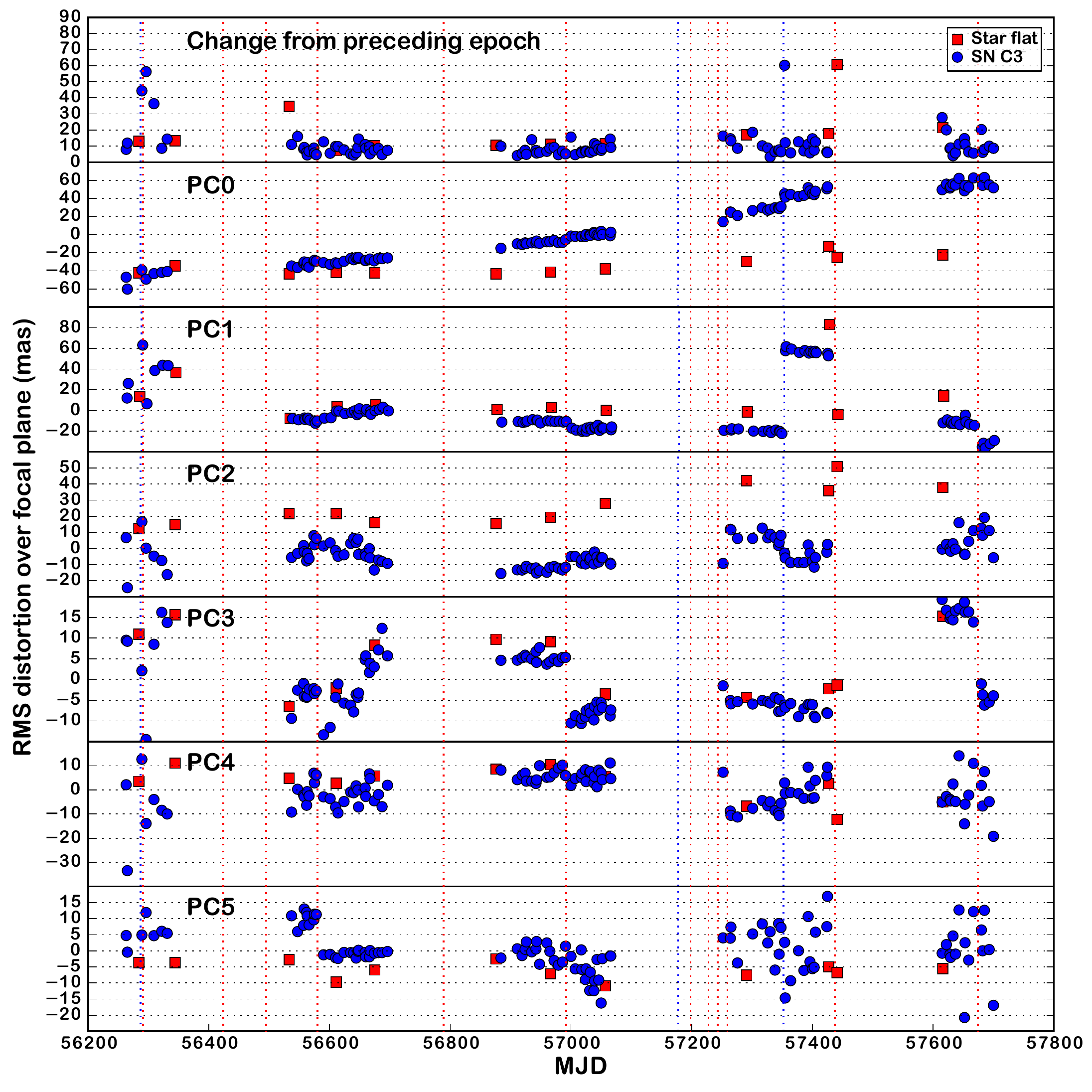}
\caption[]{\small The top row shows the RMS change across the FOV in
  the CCD shift solution between each epoch and its predecessor,
  plotted against date of observation.  The
  largest changes occur for the epochs following a warming of the
  camera to ambient temperature (marked by red vertical dashed lines)
  or a cooling of the focal plane to -120~C (blue vertical lines).
  Star flat epochs are distinguished from supernova C3 observations as
per the legend.  Further rows show the RMS contributions to each
epoch's CCD shifts of the first six principal components of variation.
 See the text for further narrative.}
\label{pca}
\end{figure}

The first row of Figure~\ref{pca} plots the difference between each
epoch's CCD shifts and the preceding epoch's solution.  We quantify
this difference by giving the RMS displacement between the solutions,
averaged over the active regions of the array.  We see immediately
that the largest changes occur for the first epoch to follow a warming
or cooling event for the camera.  DECam is cycled to room temperature
for various maintenance purposes, or when electrical power is lost for
long periods.  On three occasions the focal plane temperature dropped
from the normal $-100$~C to $-120$~C when power to its heaters was
lost for several hours.  We will refer to these as ``camera events''
and the periods between them as ``camera intervals.''

We perform a principal components analysis of the 
\edit{354}-element feature vector specifying each exposure's CCD shifts, in
hopes of revealing the temporal structure of the largest contributors
to astrometric variation.  We should be aware, however, that two
spurious signals will be present in these data:
\begin{enumerate}
\item The SNC3 exposures are taken with minimal dithering, and have
  only $\approx20$ high-$S/N$ stars per CCD, many fewer than the star flat data.  This
  means that the CCD shift fits will be pulled by the proper motions of
  the stars within each device.  We should therefore expect to see one
  or more principal components (PCs) containing a signature that is linear
  in time for the SNC3 exposures and absent from the star flat
  exposures.  Parallax motion of the SNC3 stars should be a small
  perturbation to this.
\item The SNC3 CCD shifts have a different reference epoch than the
  star flat solution's.  Therefore we should see a static difference
  between these two in at least PC.
\end{enumerate}

Figure~\ref{pca} presents the coefficients of the \edit{six} most
significant PCs, plotted against date of the epoch solution.  These
are again plotted in units of the RMS displacement they represent on
the focal plane.  We immediately identify PC0, the largest
contributor, as the expected signature of proper motion in the SNC3
stars (although it is also possible that a smaller, linear-with-time
focal plane drift is also in this PC).  We also suspect that PC2
contains the second expected spurious signal, the distinct reference
epochs for the SNC3 and star flat solutions.

\begin{figure}[t]
\center
\includegraphics[width=0.6\columnwidth]{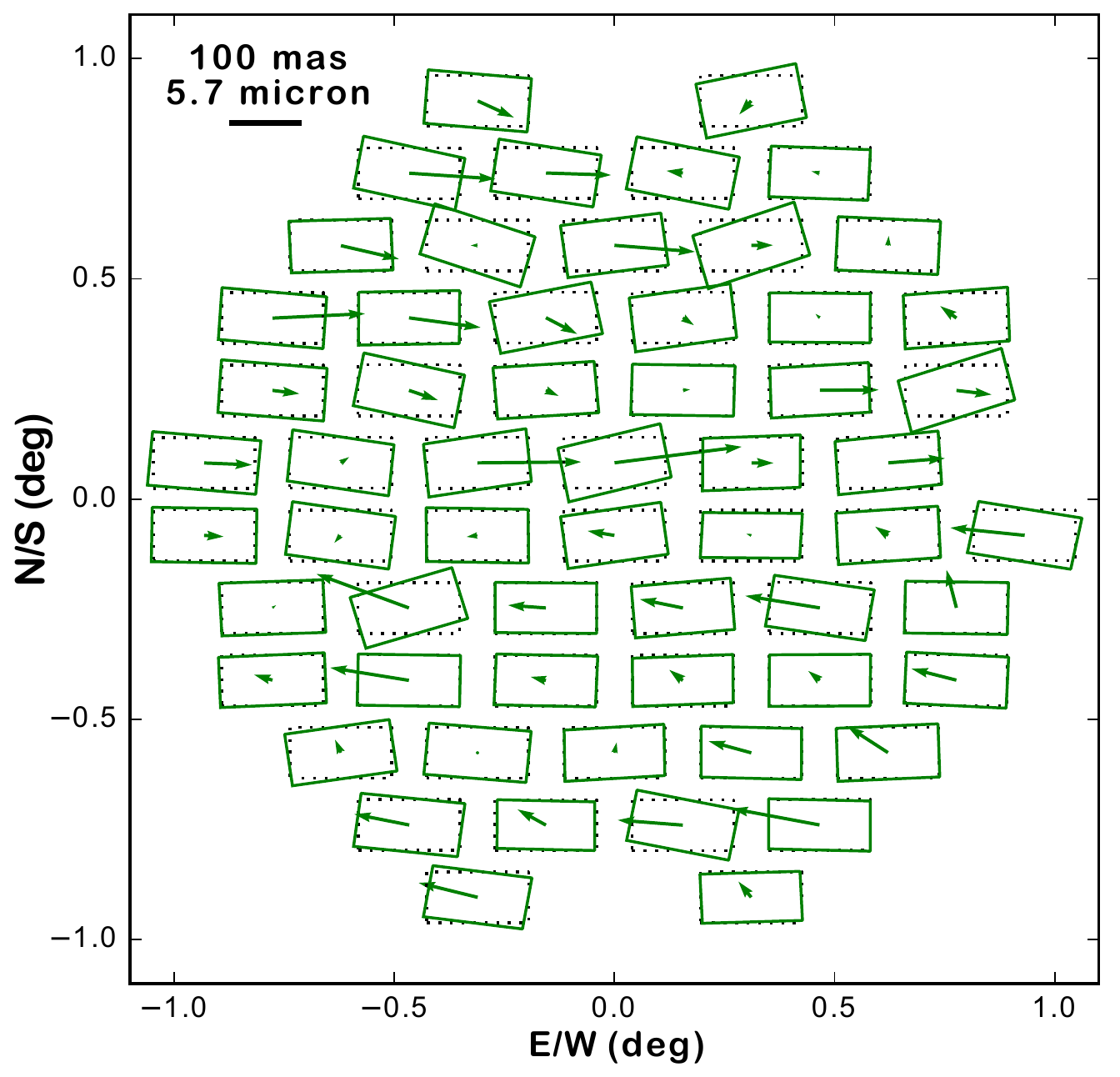}
\caption[]{\small The CCD shifts that occurred during the focal plane
  cooling of 26 Nov 2015, \ie\ PC1.  Devices translate as much as
  200~mas, nearly a full pixel, and undergo substantial rotation.
  Shear and magnification are small ($<3$~mas RMS), as expected if the
  shifts are due to displacements of the CCD carriers.}
\label{cooldown}
\end{figure}

Most striking is that the other PCs appear to be dominated by changes
that occur at camera events.  The largest, PC1, began on Thanksgiving
Day 2015, and its associated distortion persisted at normal operating
temperatures until the camera was
warmed on 19--23 Feb 2016.  Note that this warmup occurs between the last
two star flat epochs plotted in Figure~\ref{sfdrift}.
Figure~\ref{cooldown} plots the change in focal plane mapping that
occurred during this cooldown.  Some of the CCDs appear to have
moved by up to 100--150~mas, or 6--10~$\mu$m.  CCDs also show
significant rotations, contributing $\approx20$~mas RMS displacement.
Scale changes or shears of the CCDs are much smaller ($\lesssim3$~mas
RMS), as expected if the cooldown distorted the mounting structure.

Concluding that \emph{the bulk of the change in astrometric maps
  occurs during camera events,} we \edit{adopt a scheme for astrometric
calibration of DES data whereby \textbf{
each exposure is assumed to have the same instrumental astrometric map as the
nearest-in-time star flat epoch that lies within the same camera
interval.}}

\begin{figure}[t]
\center
\includegraphics[width=0.8\columnwidth]{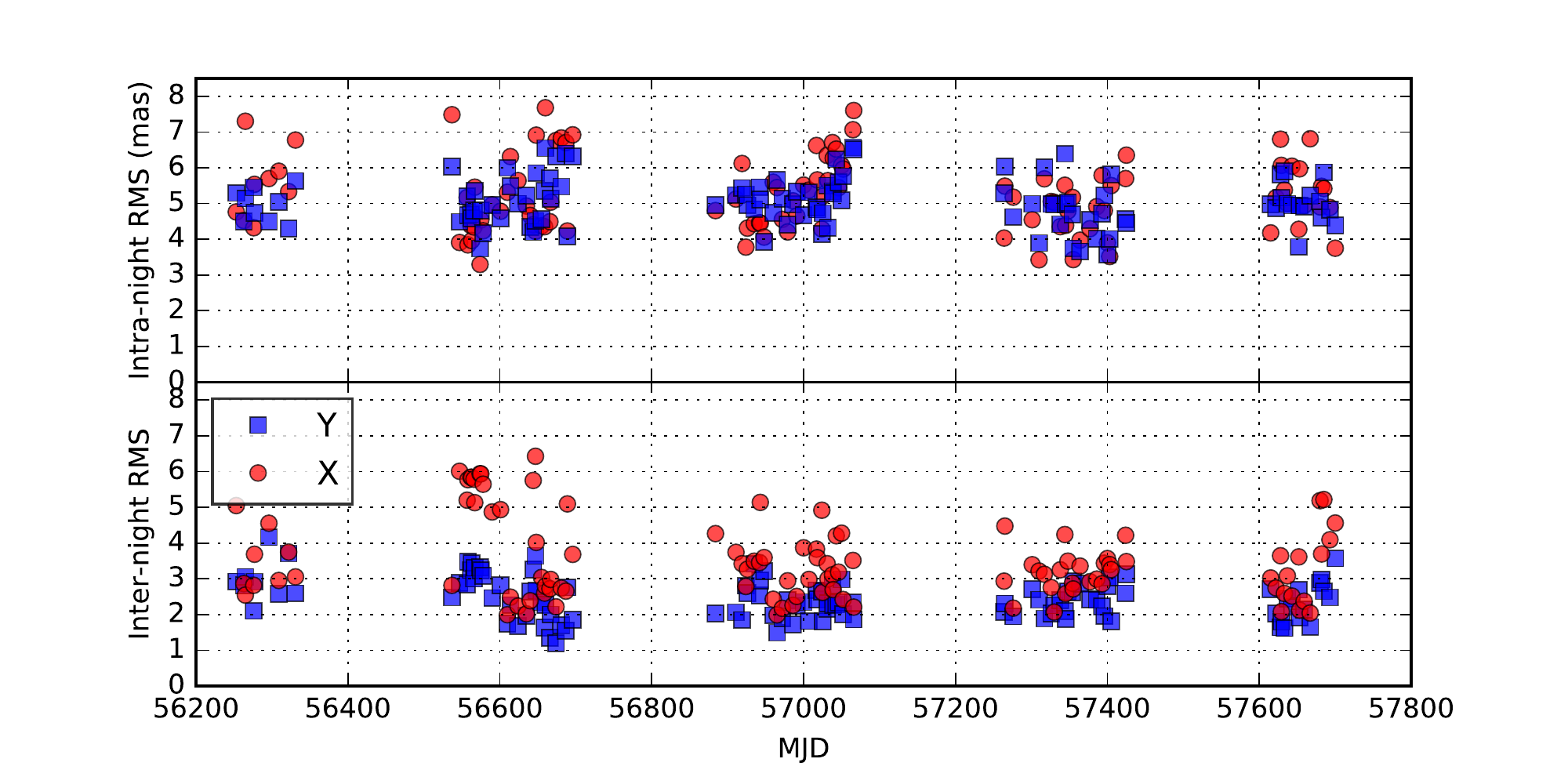}
\caption[]{\small The RMS errors of each night's SNC3 stellar
  positions are plotted 
  after each exposure has been modelled with CCD shifts taken from the
  temporally nearest star-flat image, and a free cubic polynomial
  distortion across the focal plane.  A linear proper motion has been
  fit to and removed from each star's measurements.
The upper panel plots the
  RMS residual of single-exposure positions relative to the mean
  position of the star on that night, \ie\ it gives the amplitude of
  stochastic atmospheric effects or other errors accrued between the
  10--20\arcsec\ dithers of the SN exposures.  The lower panel plots
  the RMS residual of the nightly average position against the mean of
  the entire survey.  The RMS expected from the intra-night errors has
  also been subtracted, leaving an estimate of astrometric errors that
  are coherent during a night.  These are seen to be 2--4~mas RMS,
  with a tendency to be larger in the E-W direction (red) than N-S (blue).
}
\label{snresid}
\end{figure}

\edit{We test this DES astrometric procedure by re-running \wcsfit\ on
  the SNC3 $iz$-band
  exposures using a model with the CCD shifts assigned per the local
  star flat data, and allowing each exposure a free cubic polynomial
  distortion across the focal plane. \footnote{A small number of SNC3 epochs do not have any star
  flats taken in the same camera interval; these are not
  included in Figure~\ref{snresid}.}  We fit a linear
  function of time to each SNC3 star's measured positions and subtract
  these to yield measurement errors free of proper motion.}

\edit{Figure~\ref{snresid} illustrates the quality of the resulting fits.
  In the top row we plot the RMS deviation of all stellar residuals
  from a given night relative to the mean \emph{within} the night.
  This is a measure of astrometric map components that vary on time
  scales of hours or less, which we expect to be dominated by
  atmospheric effects.  At 5--7~mas RMS per component, these are
  roughly consistent with the values seen in Figure~\ref{astrohist}
  after considering the $\sqrt{330/30}\approx3.3\times$ reduction
  in atmospheric noise expected from the longer exposure times in the
  SN field.  It is also possible that some of this RMS is due to
  unmodelled instrumental effects that vary across the
  10--20\arcsec\ dithers of the supernova exposures, \eg\ the
  pixel-to-pixel fluctuations in CCD gate lithography.}

\edit{The lower panel of Figure~\ref{snresid} plots the RMS deviation of
  each night's stellar exposures after we \emph{average the night's
    measurements of each star and subtract the noise expected from the
    intra-night variations.}  This yields an estimate of the RMS
  astrometric error that is coherent through a night, such as might be
  attributable to unmodelled shifts in the CCD positions or changes in
  optical alignment.  Such errors are seen to be in the 2--4~mas range
  on most nights.  The errors appear larger in the E-W direction than
  the N-S direction, particularly during early Y1 observations when the E-W
  errors reach 6~mas RMS, perhaps indicating lower quality in the star
  flat solution derived for the corresponding camera interval.
}


We note, finally, that the cubic polynomials we fit to each exposure
are often much larger in amplitude than could be ascribed
to atmospheric effects.  This suggests that changes in optical
alignment over time are significant at $\sim100$~mas level.

\section{Interpolation schemes}
\label{interpolation}
Given an astrometric reference catalog with errors at mas scale and
$>1$ star in each 5--10\arcmin\ coherence patch, one could measure
some fraction of the atmospheric (or other) astrometric errors and add
them to the solution, \ie\ interpolate the map between reference
stars.  The Gaia catalog will provide such a reference catalog.  The
Gaia DR1 secondary catalog \citep{gaia} does not contain proper
motions so falls slightly short of our ideal, but these will appear in
the DR2 release schedule for April
2018.\footnote{\url{http://www.cosmos.esa.int/web/gaia/release}}

Reference stars could also be obtained by repeated ground-based
observations to average atmospheric and instrumental effects.  We use
this approach for a cursory investigation of the potential of
reference-catalog interpolation.  We extract as a reference catalog
the mean positions over all star flat observations of a randomly
selected set of stars with mean density of 0.75~arcmin$^{-2}$.  These
``truth'' positions are used to interpolate the astrometric
distortions for individual exposures.  Because our star flat
observations span multiple years, the truth positions may also be
degraded by proper motions.

We use the \textsc{scikit-learn} implementation of Gaussian process
(GP) regression to interpolate the errors in the astrometric model on
a given exposure.  The GP technique requires a kernel specifying the
covariance between the error vectors of two stars separated by \vx.
We take this covariance function to have a white-noise
($\delta$-function) component of amplitude $(4\,\textrm{mas})^2$ plus
a Gaussian with amplitude $(3\,\textrm{mas})^2$ in the cross-wind
direction and $(15\,\textrm{mas})^2$ in the wind direction.  The
cross- and along-wind components of \vx\ have independent GP models. The
procedure is to:
\begin{enumerate}
\item Randomly select a training set of stars at the chosen density
  and fit the GP model to these.
\item Use the GP to interpolate to the location of each training star,
  and reject training stars with outlier residuals (\eg\
  high-proper-motion stars).  
\item Refit the GP using the retained training stars.
\item Interpolate to the positions the validation set of remaining high-$S/N$ stellar
  detections.
\item Remove outlying residuals from the validation set.
\item Calculate the 2-point correlation functions of the residuals.
\end{enumerate}
\begin{figure}[t]
\center
\includegraphics[width=0.7\columnwidth]{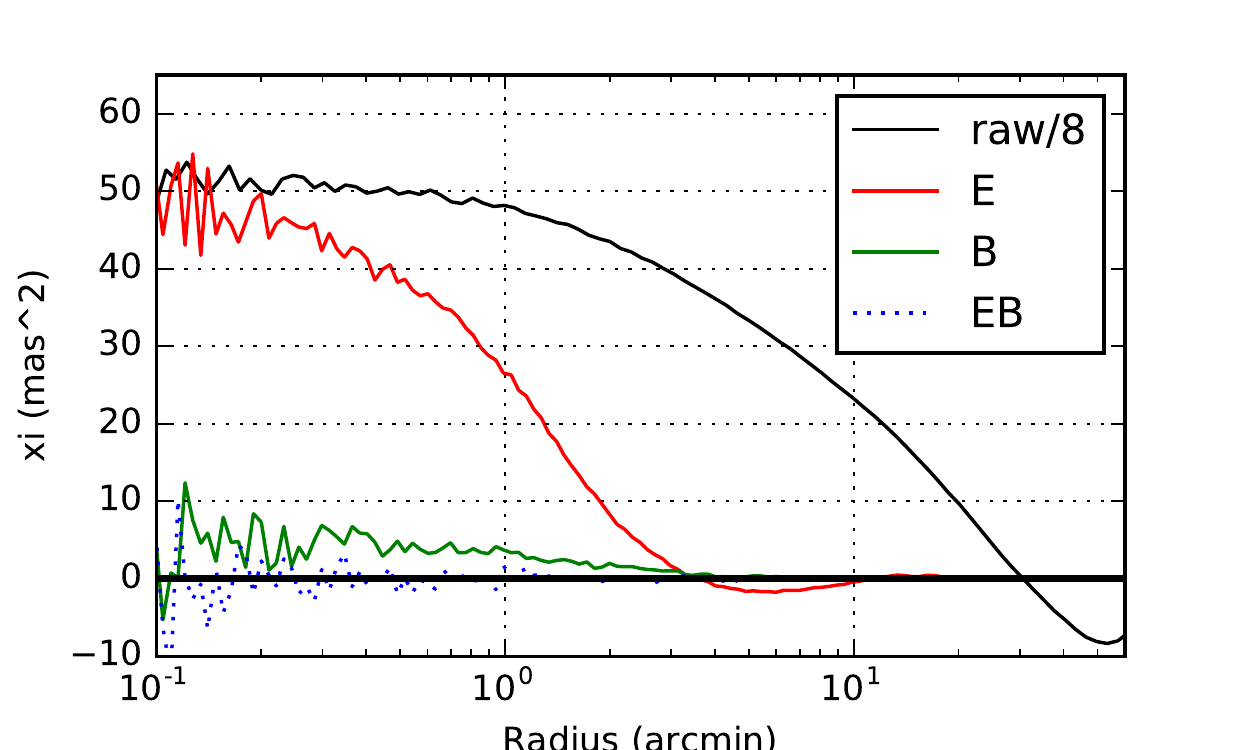}
\caption[]{\small The 2-point correlation functions $\xi_E, \xi_B,$
  and $\xi_\times$ are plotted vs separation both before (``raw'')
  and after interpolation of the astrometric errors
  using a reference star set of density 0.75~arcmin$^{-2}$.  The plot
  shows the mean $\xi$ across $21\times30$ s $z$-band exposures in star-flat epoch
  \texttt{20130829.}  Note that we plot only $\xi_E$ for the
  pre-interpolation case (black) since we have found the distortions
  consistent with pure $E$-mode behavior, and that the
  pre-interpolation plot is reduced by a factor of 8 to fit on the
  same plot.  The reduction in $\xi$ from interpolation is dramatic,
  with correlations at scales above the reference-star density being
  essentially eliminated.}
\label{postinterp}
\end{figure}

We execute this process for 21 $z$-band exposures in epoch
\texttt{20130829}, the same set plotted in Figure~\ref{ebcorr}.  The
mean correlation functions before and after interpolation are plotted
in Figure~\ref{postinterp}.  The $E$ mode remains dominant even though
the interpolation process is not designed to conserve $E/B$ behavior.
As expected, the interpolation reduces $\xi_E$ to negligible levels
($\lesssim1$~mas$^2$) at scales $>3\arcmin$ where multiple reference
stars can contribute to interpolation.  The removal of large-scale
power reduces the $\xi_E(r)$ at $r\rightarrow0$ by a factor $\approx
8$ from the pre-interpolation value.  The average post-interpolation residuals
are $<7$~mas RMS for this epoch, which has typical stochastic signal
level $\xi_0$.  The correlation length of the astrometric errors is
reduced to 1\arcmin.  One would expect the amplitude of the
post-interpolation residuals to decrease with the square root of
integration time until the systematic error floor of either the
reference catalog or DECam is reached.

This is just an initial investigation: we have attempted to
optimize the interpolation procedure neither for accuracy nor speed.
Certainly there is improvement to be had through GP kernel
optimization or other approaches, including interpolation schemes that
exploit the known absence of $B$ modes in the vector distortion field.
Doubling the density of reference stars appears to have little effect
on the residual amplitude.

\section{Conclusions}
\label{conclusion}

An astrometric model for DECam with errors near mas level requires
terms not only for the polynomial optical ``plate solution,'' but also
contributions from: stray electric fields near the edges in the
detector and from ``tree ring'' impurity fluctuations; lateral color
and differential chromatic refraction in the bluer bands; shifts
in the CCD positions primarily accrued during focal-plane temperature
excursions; and time-variable low-order (cubic) distortions across the
FOV from a litany of instrumental and atmospheric effects.

All of these distortions components are determined reliably by fitting
a model to stellar positions measured from dithered DECam exposures.
External reference catalogs play little role in this process, being
needed only to stabilize some large-scale degeneracies such as the
overall pixel scale.  The \wcsfit\ software that we created for this
purpose is similar to the widely-used \scamp\ code in optimizing the
parameters of a model to maximize agreement among multiple exposures
of the same star.  \wcsfit\ uses simple linearized iterations to
minimize a $\chi^2$, relying on \scamp\ or some other code to have
produced an initial solution that maps each exposure to
$\lesssim1\arcsec$ accuracy.  \wcsfit\ complements \scamp\ by:
the ability to specify and fit a complex model with many components
interlacing their effects among many exposures; enhanced outlier
rejection, necessary to achieve precise modelling; and some
optimizations for fitting large exposure sets with large numbers of
free parameters.

Once this model is fit to an ensemble of DECam exposures, the
remaining astrometric errors are dominated by a curl-free stochastic
field of atmospheric refraction fluctuations.  In a typical 30 s
exposure, the stochastic atmospheric distortions are 10--30~mas RMS
with coherence length of 4-10\arcmin\ and a strongly anisotropic
pattern from wind-blown turbulence.  Some nights are much worse than
this; unfortunately there is no strong connection between seeing FWHM
and astrometric quality.

The atmospheric turbulence averages down with longer exposures or
through stacking of residuals on many exposures.  Doing so reveals
weaker but persistent errors in the astrometric model.  
Fixed patterns in the devices at 2--4~mas RMS (0.008-0.015 pixel)
are dominated by larger residuals on small regions of the device subject
to edge effects and mounting structures.  These could be tabulated
from the device stacks and added to the model if we acquired even more
stellar measurements.

Star-flat exposures sequences taken every few months are used to
monitor shifts in CCD positions.  Using DES supernova-field
observations, we determine that the bulk of the observed shifts occurs
when DECam's focal plan warms to room temperature or cools below
normal operating temperature.  If we apply the CCD shifts measured in
the star-flat epochs to the SN data, we find that remaining errors
inter-night variation in the solution is 2--4~mas.

The 4\arcmin--10\arcmin\ coherence length of the dominant atmospheric
distortions suggests that the Gaia reference catalog, with positions
and (in the future) proper motions for
$\approx1$ star per arcmin$^2$ at high Galactic latitude, can be used
to constrain and remove the atmospheric pattern (and, trivially, the
low-order polynomial distortions).  Indeed we find that a trial of
Gaussian-process interpolation using reference stars at this density
reduces the correlation function $\xi_+(r)$ of errors to $<1$~mas$^2$
on scales $r>3\arcmin$ and reduce the RMS value at smaller scales to
$<7$~mas in a 30 s exposure.

\edit{The astrometric maps and procedures derived herein will be
  applied to the DES observations and made available to other users of
  DECam.  We find that errors in the resultant positions are likely to
  be dominated by (in order of decreasing importance):}

\edit{\begin{itemize}
\item Unavoidable shot noise in the measurement of object centers (for
  fainter stars).
\item Atmospheric turbulence of typical RMS amplitude (15--20)$\times
  \sqrt{30/T}$~mas, where $T$ is the exposure time in seconds, and
  coherence length $\approx10\arcmin.$  This
  amplitude depends on the weather.  Given a Gaia catalog with proper
  motions, the atmospheric field can be measured and interpolated to
  random locations, leaving residuals $\approx 3\times$ smaller and
  with coherence length $\approx1\arcmin.$
\item Unmodelled night-to-night variations in the DECam astrometric
  solution at 2--4~mas RMS.
\item Up to 2--4~mas RMS additional errors from static detector
  effects (mounting holes, edge fields) that we do not yet model, but
  could potentially include given a much larger set of stellar
  measurement residuals.
\end{itemize}
}

We conclude that the DECam astrometric model, with registration to the
Gaia catalog, has RMS errors below 10~mas in a typical 30 s
exposure, small enough to be negligible for cosmic-shear measurements,
and likely to be even smaller in the standard 90-second DES exposure.
For a general-use, wide-field instrument like DECam to reach the limit
of astrometric accuracy imposed by atmospheric
turbulence (with Gaia interpolation gains), the best
observing scheme is to dither successive exposures so that the few mas
of remaining systematic camera-centered distortions are sampled
differently for each exposure of the desired targets.  Such a strategy
is intrinsic to the DES Wide 5000~deg$^2$ survey, so we should expect
astrometric catalogs from this survey that are limited by the
combination of image shot noise and atmospheric turbulence.

The LSST aims to achieve this goal as well, and the DECam results here
show that this should be entirely feasible.  LSST has a larger field,
shorter exposures, and many more stellar detections to work
with.  The DECam experience perhaps shows the value of regular
star-flat observation sequences.  A substantial complication for LSST
(as well as other modern wide-field imagers/telescopes such as Hyper
Suprime-Cam) is its alt-az mounting and consequent need of an
instrument rotator.  This introduces a degree of freedom to the
optical system absent from the equatorial-mounted Blanco telescope,
perhaps greatly increasing the number of constraints that must be
analysed to yield a solution valid at all rotator angles.  LSST will
obtain many more stellar images, so the necessary data will likely
exist but pose a bigger computational challenge.

\acknowledgements
GMB gratefully acknowledges support from grants AST-1311924 and AST-1615555
from the National Science Foundation, and DE-SC0007901 from the Department
of Energy. AAP is supported by the Jet Propulsion Laboratory. Part of
the research was carried out at the Jet Propulsion Laboratory,
California Institute of Technology, under a contract with the National
Aeronautics and Space Administration.  We thank Steve Holland and Greg
Derylo for help with interpretations of detector and camera behaviors.

The \wcsfit\ code and subsequent analyses make extensive use of the
following excellent public software packages: \textsc{tmv} for linear
algebra and \textsc{TreeCorr} for fast correlation functions, both by
R. M. Jarvis; \textsc{yaml-cpp} by J. Beder for YAML decoding; and
\textsc{cfitsio} by W. Pence, \textsc{fitsio} by E. Sheldon, and
\textsc{AstroPy} for FITS access in C and Python environments.

Funding for the DES Projects has been provided by the U.S. Department of Energy, the U.S. National Science Foundation, the Ministry of Science and Education of Spain, 
the Science and Technology Facilities Council of the United Kingdom, the Higher Education Funding Council for England, the National Center for Supercomputing 
Applications at the University of Illinois at Urbana-Champaign, the Kavli Institute of Cosmological Physics at the University of Chicago, 
the Center for Cosmology and Astro-Particle Physics at the Ohio State University,
the Mitchell Institute for Fundamental Physics and Astronomy at Texas A\&M University, Financiadora de Estudos e Projetos, 
Funda{\c c}{\~a}o Carlos Chagas Filho de Amparo {\`a} Pesquisa do Estado do Rio de Janeiro, Conselho Nacional de Desenvolvimento Cient{\'i}fico e Tecnol{\'o}gico and 
the Minist{\'e}rio da Ci{\^e}ncia, Tecnologia e Inova{\c c}{\~a}o, the Deutsche Forschungsgemeinschaft and the Collaborating Institutions in the Dark Energy Survey. 

The Collaborating Institutions are Argonne National Laboratory, the University of California at Santa Cruz, the University of Cambridge, Centro de Investigaciones Energ{\'e}ticas, 
Medioambientales y Tecnol{\'o}gicas-Madrid, the University of Chicago, University College London, the DES-Brazil Consortium, the University of Edinburgh, 
the Eidgen{\"o}ssische Technische Hochschule (ETH) Z{\"u}rich, 
Fermi National Accelerator Laboratory, the University of Illinois at Urbana-Champaign, the Institut de Ci{\`e}ncies de l'Espai (IEEC/CSIC), 
the Institut de F{\'i}sica d'Altes Energies, Lawrence Berkeley National Laboratory, the Ludwig-Maximilians Universit{\"a}t M{\"u}nchen and the associated Excellence Cluster Universe, 
the University of Michigan, the National Optical Astronomy Observatory, the University of Nottingham, The Ohio State University, the University of Pennsylvania, the University of Portsmouth, 
SLAC National Accelerator Laboratory, Stanford University, the University of Sussex, Texas A\&M University, and the OzDES Membership Consortium.

The DES data management system is supported by the National Science Foundation under Grant Number AST-1138766.
The DES participants from Spanish institutions are partially supported
by MINECO under grants AYA2015-71825, ESP2015-88861, FPA2015-68048,
SEV-2012-0234, SEV-2012-0249, and MDM-2015-0509, some of which include
ERDF funds from the European Union. IFAE is partially funded by the
CERCA program of the Generalitat de Catalunya.

\appendix
\section{E/B vector field correlation functions}
\label{ebappendix}
We wish to calculate the correlation functions of the curl-free and
divergence-free components of the residual astrometric distortion
field $\Delta\vx$ on each exposure, given an irregular, noisy sampling of this
field by stellar detections.  This is closely analagous to the E/B
decompositions performed on the spin-2 polarization field of the
cosmic microwave background and the spin-2 weak gravitational shear
field in many cosmological investigations.  We can derive the vector
E/B decomposition by a very slight alteration to the shear-field
derivation given by \citet{SvWM}.

We start with a 2d vector field $\vv=(v_x, v_y)$.  It is useful to
work with a complex notation $v = v_x + iv_y$ and complex derivatives
$\partial = \partial_x + i\partial_y.$  We can write an arbitrary
vector field as
\begin{align}
\label{ebderiv}
v & = \partial \phi(\vx), \\
\phi & \equiv \phi_E + i\phi_B.
\end{align}
The curl-free E mode of \vv\ is sourced by $\phi_E$ and the
divergence-free B mode by $\phi_B.$

The 2-point correlation functions of \vv\ at separation vector \vvr\
are best posed in terms of the quantity $v_\parallel + i v_\perp = v
e^{-i\beta}$, where $\beta$ is the position angle of \vvr.  We define
\begin{align}
\label{xiplus}
\xi_+(\vvr) = \xi_+(r,\beta) & = \left\langle v(\vx) e^{-i\beta}
                              \left[v(\vx+\vvr)
                              e^{-i\beta}\right]^\star \right\rangle 
  \\
\label{ximinus}
\xi_-(r,\beta) + i\xi_\times(r,\beta)  & = \left\langle v(\vx) e^{-i\beta}
                              v(\vx+\vvr)
                              e^{-i\beta} \right\rangle
\end{align}

Taking $\tilde \phi(\vk)$ to be the Fourier transform of $\phi$, and
the generation of $\phi$ to be a stationary stochastic process, we
define the power spectra via
\begin{align}
\label{ebpower}
\left\langle \tilde \phi_E(\vk_1) \tilde \phi^\star_E(\vk_2)
  \right\rangle & = (2\pi)^2 \delta(\vk_1-\vk_2) P_{EE}(\vk_1) \\
\nonumber
\left\langle \tilde \phi_E(\vk_1) \tilde \phi^\star_B(\vk_2)
  \right\rangle & = (2\pi)^2 \delta(\vk_1-\vk_2) P_{EB}(\vk_1) \\
\nonumber
\left\langle \tilde \phi_B(\vk_1) \tilde \phi^\star_B(\vk_2)
  \right\rangle & = (2\pi)^2 \delta(\vk_1-\vk_2) P_{BB}(\vk_1).
\end{align}
$P_{EB}$ must be real if the vector field statistics are invariant
under 180\arcdeg\ rotation, so we will assume this is true.  The real
part will vanish as well if the process generating \vv\ is invariant
under parity flips.  We will leave the real part as a free parameter.

By propagating the derivatives in \eqq{ebderiv} through a Fourier
transform we can express the correlation functions (\ref{xiplus}) and
(\ref{ximinus}) as
\begin{align}
\xi_+(r,\beta) & = \int d\alpha \int k\, dk \, \left|
                 k^2 \right| e^{ikr \cos(\alpha-\beta)} \left[
                 P_{EE}(k,\alpha) + P_{BB}(k,\alpha)\right] \\
\xi_-(r,\beta) + i \xi_\times(r,\beta) & = \int d\alpha \int k\, dk \, \left|
                 k^2 \right| e^{-ikr \cos(\alpha-\beta)}
                                         e^{2i(\alpha-\beta)} \left[
                 P_{EE}(k,\alpha) - P_{BB}(k,\alpha) + 2i
                                         P_{EB}(k,\alpha)\right]
\end{align}
where $\alpha$ is the position angle of the wavevector \vk.  Even
though the atmospheric distortions are anisotropic, we will concern
ourselves only with the angle-average quantities $\xi_+(r)=\langle
\xi_+(r,\beta) \rangle_\beta,$ and the corresponding angle-averaged
power spectra $P_{EE}(k),$ etc.  If we average the preceding equations
over $\beta$, Bessel's first integral yields
\begin{align}
\label{xip1}
\xi_+(r) & = 2\pi \int k\, dk \,    k^2  J_0(kr) \left[
                 P_{EE}(k) + P_{BB}(k)\right] \\
\label{xim1}
\xi_-(r)  & = -2\pi \int k\, dk \, k^2 J_2(kr)
                 \left[ P_{EE}(k) - P_{BB}(k)\right] \\
\label{xix}
\xi_\times(r) & = -4\pi \int k\, dk \, k^2 J_2(kr) P_{EB}(k).
\end{align}
The last equation tells us that $\xi_\times$ is produced purely by EB
power.  Let us define pure-E and pure-B quantities
\begin{align}
\label{xiEB}
\xi_{E,B}(r) & \equiv 2 \pi \int k\, dk\, k^2 J_0(kr) P_{EE,BB}(k) \\
 & = \frac{1}{2} \left\{ \xi_+(r)  \mp \int r\,dr\, J_0(r) \int
   r^\prime \, dr^\prime J_2(r^\prime) \xi_-(r^\prime). \right\}
\end{align}
where the last line combines \eqq{xip1} with the order-2 Hankel
Transform of \eqq{xim1}.  After making use of identities 9.1.27 and
11.4.42 from \citet{AS}, this can be converted to
\begin{equation}
\label{xiEB2}
\xi_{E,B}(r) = \frac{1}{2} \left\{ \xi_+(r)  \pm \left[
    \xi_-(r) - 2 \int_r^\infty dr^\prime \frac{1}{r^\prime} \xi_-(r^\prime)\right] \right\}.
\end{equation}

\begin{figure}
\center
\includegraphics[width=0.6\columnwidth]{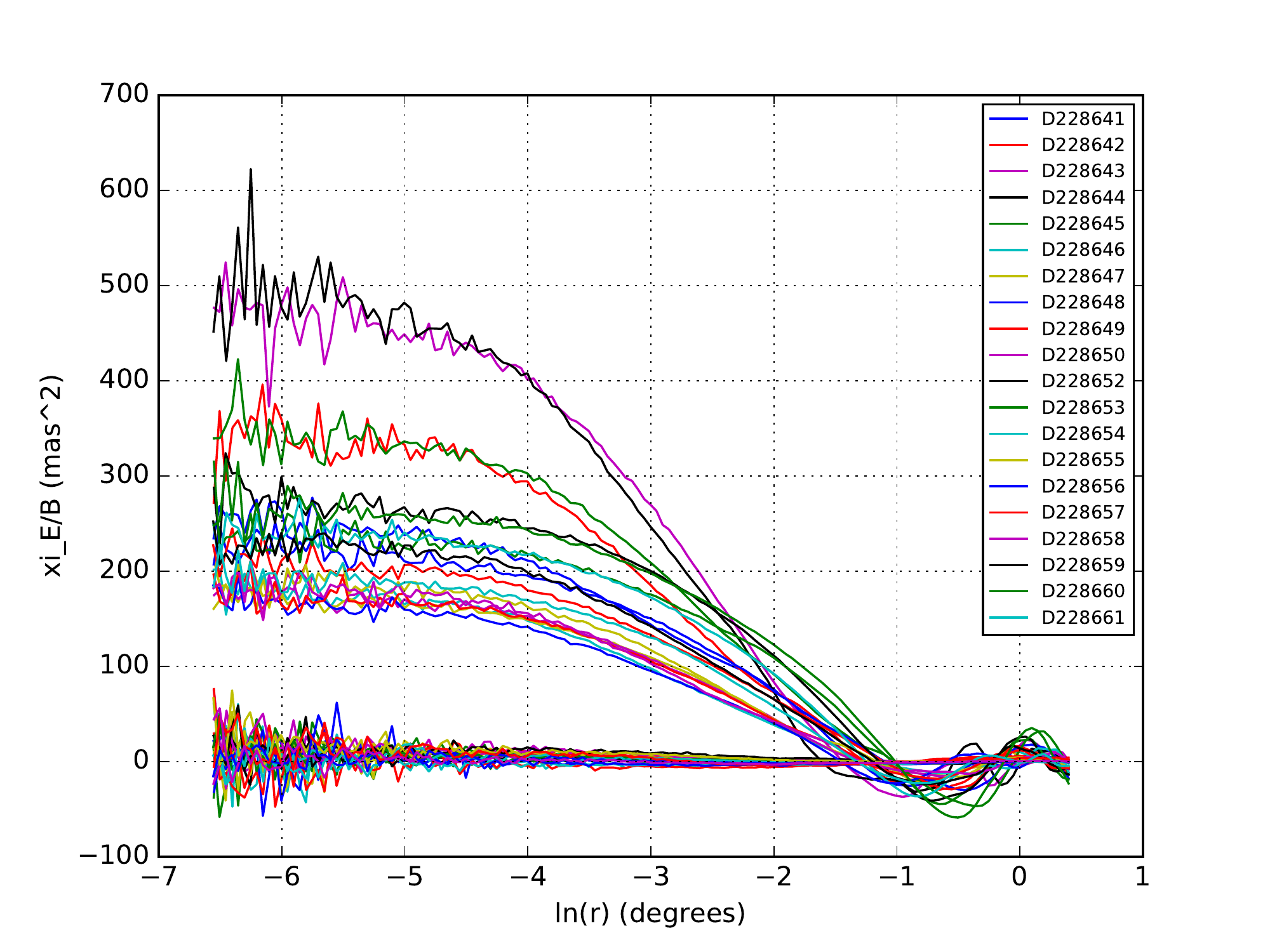}
\caption[]{\small The $\xi_E$ (upper curves) and $\xi_B$ (lower set)
  derived via \eqq{xiEB2} for each of a series of exposures,
  demonstrating that the astrometric errors are dominated by a
  curl-free vector field, as expected from atmospheric refraction
  fluctuations.
}
\label{ebcurves}
\end{figure}

Equations~(\ref{xix}) and (\ref{xiEB2}) allow us to produce 
measures of pure E, B, and cross-EB power from 2-point correlations
constructed from all pairs of detections in a given exposure.  In
Figure~\ref{ebcurves} we plot $\xi_E$ and $\xi_B$ inferred for the
astrometric residuals in each of 20
consecutive star-flat exposures (after projecting out a cubic
polynomial function of field coordinates from each exposure).  It is
clear that the astrometric residuals are indeed dominated by E modes,
\eg\ curl-free.  Figure~\ref{ebcorr} plots the mean of $\xi_E, \xi_B,$ and
$\xi_\times$ in another set of exposures, confirming that any
divergence modes are very small.  In further analysis we assumed
$\xi_B=0$ such that we can more simply take $\xi_E=\xi_+$.

\end{document}